\documentclass[twocolumn,superscriptaddress]{revtex4-2}
\usepackage{gnuplottex}
\usepackage{mathtools}
\usepackage{float}
\usepackage{array}
\usepackage[english]{babel}
\usepackage[utf8]{inputenc}
\usepackage{amsmath}
\usepackage[table]{xcolor}
\usepackage{amsfonts}
\usepackage{graphicx}
\usepackage{dsfont}
\usepackage[colorlinks=true, linkcolor=blue, citecolor=dred]{hyperref}
\usepackage{algorithm}
\usepackage[noend]{algpseudocode}
\usepackage{bbold}
\usepackage{mathtools}
\usepackage{amssymb}
\usepackage{bm}
\usepackage{bbm}
\usepackage{tabularx, booktabs}
\usepackage{xspace}
\usepackage{listings}
\usepackage[noend]{algpseudocode}
\newcolumntype{Y}{>{\centering\arraybackslash}X}
\definecolor{dgreen}{rgb}{0,.5,0}
\definecolor{dblue}{rgb}{0,0,.5}
\definecolor{dred}{rgb}{0.5,0,.5}

\usepackage{physics}

\newcommand\reallywidehat[1]{%
\savestack{\tmpbox}{\stretchto{%
  \scaleto{%
    \scalerel*[\widthof{\ensuremath{#1}}]{\kern-.6pt\bigwedge\kern-.6pt}%
    {\rule[-\textheight/2]{1ex}{\textheight}}%WIDTH-LIMITED BIG WEDGE
  }{\textheight}% 
}{0.5ex}}%
\stackon[1pt]{#1}{\tmpbox}%
}
\begin{document}

% \title{The extended star graph as a light-harvesting-complex prototype: \\ impact of decoherence on the excitonic absorption speedup mediated by local defects  
% } % Article title
% \title{ Extended star graph as a light-harvesting-complex prototype:  impact of a dephasing environment on the excitonic absorption speedup mediated by energy defects \\
% \saad{Titre plus general}
% } 

\title{ 
% Effect of a dephasing environment on the excitonic absorption speedup by energy defect tuning: arising of environment assisted quantum transport
% \\
% Excitonic absorption speedup by energy defect tuning: effect of a dephasing environment in complex networks 
% \\
% Excitonic absorption in extended and linear networks: \\ interplay between optimally tuned energy defects and dephasing environment 
} 

 \title{  
 Optimized excitonic transport mediated by local energy defects:\\ survival of optimization laws in the presence of dephasing
}

% Extended star graph as a light-harvesting-complex prototype: Excitonic absorption speedup by peripheral energy defect tuning

\author{Lucie Pepe } 
\affiliation{Laboratoire de Chimie Quantique, Institut de Chimie,
CNRS/Université de Strasbourg, 4 rue Blaise Pascal, 67000 Strasbourg, France} 
\author{Vincent Pouthier} 
\affiliation{Institut UTINAM,  Universit\'{e} de Franche-Comt\'{e}, CNRS UMR 6213, 25030 Besan\c {c}on, France} 
\author{Saad Yalouz} 
\email{yalouzsaad@gmail.com}
\affiliation{Laboratoire de Chimie Quantique, Institut de Chimie,
CNRS/Université de Strasbourg, 4 rue Blaise Pascal, 67000 Strasbourg, France}

\begin{abstract}

In an extended star with peripheral defects and a core occupied by a trap, it has been shown that exciton-mediated energy transport from the periphery to the core can be optimized [S. Yalouz et al. Phys. Rev. E \textbf{106}, 064313 (2022)]. 
If the defects are judiciously chosen, the exciton dynamics is isomorphic to that of an asymmetric chain and a speedup of the excitonic propagation is observed. 
Here, we extend this previous work by considering that the exciton in both an extended star and an asymmetric chain, is perturbed by the presence of a dephasing environment. 
Simulating the dynamics using a Lindblad master equation, two questions are addressed: how does the environment affect the energy transport on these two networks? And, do the two systems still behave equivalently in the presence of dephasing? 
Our results reveal that the time-scale for the exciton dynamics strongly depends on the nature of the network.
But quite surprisingly, the two networks behave similarly regarding the survival of their optimization law. 
In both cases, the energy transport can be improved using the same original optimal tuning of energy defects as long as the dephasing remains weak. 
However, for moderate/strong dephasing, the optimization law is lost due to quantum Zeno effect.
%
%These results suggest that different networks can react in a fairly equivalent way to an environment, and that the optimal transport conditions found for closed systems can still be valid for open systems (up to a certain limit).

\end{abstract}

\maketitle

% \saad{REFERENCES :\\
% Localisation determines the optimal noise rate for
% quantum transport}

\section{Introduction}

% \saad{Je pense que ce premier paragraphe peut etre élagué et pourquoi pas connecté a celui qui suit. }

Studying exciton-mediated energy transport in molecular lattices has a long history that can be traced back to the 60-70’s~\cite{davydov71,reineker82,may00,agranovich09}. At that time, and during the following decades, the research was realized to understand the behavior of translationally invariant molecular crystals, with particular emphasis on the characterization of their optical properties~\cite{silbey76}.  
Different features were considered, i.e., the study of the exciton-phonon interaction to explain relaxation processes, optical line-shapes, and quantum diffusion~\cite{grover70,grover71,yarkony76,wang88,brown89,sonnek96}, the influence of defects to characterize specific optical response and to investigate localization phenomena~\cite{abram75,blumen84,malyshev99,vlaming09}, and the analysis of nonlinear effects to describe both the nonlinear excitonic optical response~\cite{mukamel95} and the formation of nonlinear objects such as solitons~\cite{scott92}.

Nowadays, with the development of quantum technologies, new ideas have emerged, and it has been pointed out that exploiting exciton propagation in complex networks could be used to carry either quantum information, or energy, at nanoscale~\cite{mulken11}. 
Indeed, on complex networks, the delocalization of an exciton defines a physical realization of a continuous time quantum walk (CTQW)~\cite{mulken06}. % which has several applications in quantum information. 
Widely studied in recent years, the concept of CTQW has been used to answer a variety of questions in quantum information theory ranging from the realization of perfect quantum state transfer~\cite{christandl04} to the development of high-performance algorithms~\cite{childs09,venegas12,pouthier15}. 
On another note, this paradigm has been also widely used to study the energy transfer on complex networks with a specific focus on the role of the topology~\cite{razolli21,galiceanu16}, the effect of disorder~\cite{yalouz20,mulken11}, and the presence of traps~\cite{agliari10,agliari11} to cite but a few.
In the present paper, we are interested in this second aspect where complex networks architectures are considered as a medium for efficient exciton-mediated energy transport at the nanoscale.
Such an idea was first pointed out by Mukamel~\cite{mukamel97} 
% \lucie{Such an idea was launched by Mukamel ?}
who suggested that excitons could be exploited in dendrimers to design artificial light-harvesting complexes (LHC)~\cite{tretiak98,harigaya99,poliakov99,minami00,nakano00,kirkwood01,martin02,nakano04,supritz05,crabtree07,nakano09,pouthier14}. 
A dendrimer is a chemical tree-like structure formed by several dendritic branches that emanate out from a central core~\cite{bosman99,vogtle09,astruc10}. Therefore, the functionalization of the terminal groups by chromophores favors light harvesting.
The absorbed light generates local photo-excitations, i.e.~Frenkel excitons, that propagate along the branches and converge towards the central core which contains either a fluorescent trap, a reaction center, or a chemical sensor~\cite{bar-haim97,choi02}.

\textcolor{black}{Inspired by these tree-like architectures, the excitonic CTQW on an extended star was studied recently to highlight the realization of an efficient photo-excitation transfer~\cite{yalouz22}.
In this context, it was considered that the periphery of the star was functionalized by tunable energy defects, while the core was occupied by a trap. 
The absorption of light by the defects generates an exciton in an initial state uniformly delocalized over the peripheral sites.
The energy is then transferred via an excitonic CTQW from the periphery to the core where an irreversible absorption process occurs due to the presence of the trap.
The investigations realized on this prototype LHC allowed to evidence the possibility to strongly optimize the resulting energy transfer/absorption process.
Depending on both the number and the length of the branches, it has been shown that if the energy defects are judiciously chosen, the initial state localized at the periphery may hybridize with a state localized on the core. 
Consequently, a speedup of the excitonic propagation was observed revealing the potential of the extended star as a prototype  artificial LHC.} 

\textcolor{black}{Interestingly, in Ref.~\cite{yalouz22} it was also highlighted that the optimal excitonic transfer occurring on the extended star was actually isomorphic to that of an asymmetric chain (when the exciton is initially delocalized over the periphery of the star). 
This chain, whose length is equal to that of the branches of the star, involves sites whose meaning can be understood as follows. 
The first site refers to the trap, the second site refers to an exciton delocalized over the first site of the branches of the star, the second site refers to an exciton delocalized over the second site of the branches of the star, ... and so on. 
The resulting isomorphism between the two networks reveals a fundamental point: both the extended star and the asymmetric chain act as equivalent LHC prototypes that can be designed to produce a same optimal energy transfer/absorption. 
% All these results were obtained assuming that both quantum systems were closed, and we could now legitimately ask about the evolution of the optimal transport in the context of a more realistic open quantum system description. }
At this step, it should be noted here that all these results were obtained assuming that the two quantum systems were closed.
Naturally, we can wonder now about the evolution of the optimal transport in both systems when a more realistic description is considered, including for example the presence of an external environment.}

\textcolor{black}{
% Nevertheless, in a more realistic description of these systems, such optimal transport properties may be altered due to the presence of the external environment. 
Indeed, in nature, excitons no longer propagate freely: they behave as open quantum systems interacting with the remaining degrees of freedom of the medium,  usually associated with a phonon bath~\cite{may00}. 
The phonons are thus responsible for quantum dephasing~\cite{joos03,schlosshauer07,breuer07} that drastically modifies the way excitons delocalize. 
Generally acting as a disruptive ingredient, the presence of a phonon bath tends to prevent the conservation of superposed states. 
It thus generates a transition between an efficient coherent propagation and an inefficient incoherent diffusive motion~\cite{pouthier08,pouthier09}.
% Conversely, as pointed out by several authors~\cite{mohseni08,plenio08,rebentrost09}, note that the presence of a purely dephasing environment could improve the process of excitonic transport on a network. 
% This is the so-called Environment Assisted Quantum Transport concept in which the subtle interplay between coherent propagation and incoherent jumping processes due to the exchanges of energy with the phonons could make it possible to produce an ultra-efficient hybrid quantum transport. 
% Thus, contrary to what one might expect, quantum dephasing could also have an unsuspected positive effect.
}

\textcolor{black}{
Motivated by this view, in the present paper we want to address two questions: how would the presence of the environment affect the energy transport on both the extended star and the asymmetric chain? 
And also, do the two networks still behave equivalently in the presence of dephasing? 
To this end, the excitonic dynamics on both networks will be revisited in this work by considering the influence of an external dephasing environment.
To proceed, a standard stochastic approach is used by assuming that the environment behaves as a Gaussian Markovian $\delta$-correlated stochastic potential field acting on the exciton~\cite{haken72,haken73,jackson81,rips93,pfluegl00}.
Within this model, a generalized master equation (GME) is established for describing the time evolution of the exciton reduced density matrix (RDM). 
The knowledge of the RDM allows us to compute in principle all the observables needed to characterize the exciton dynamics.
}

% In the present paper, the excitonic dynamics is   revisited by considering the influence of an external dephasing environment. Two different tripartite systems will be considered, i.e., the extended star and the asymmetric chain. To proceed, a standard stochastic approach is used by assuming that the environment behaves as a Gaussian Markovian $\delta$-correlated stochastic potential field acting on the exciton~\cite{haken72,haken73,jackson81,rips93,pfluegl00}. Within this model, a generalized master equation (GME) is established for describing the time evolution of the exciton reduced density matrix (RDM). The knowledge of the RDM allows us to compute in principle all the observables needed for characterizing the exciton dynamics.

The paper is organized as follows, in Sec.~\ref{sec:theory} the extended star and the asymmetric chain are introduced and the exciton Hamiltonians are defined.
Then the relevant observables required for characterizing the dynamics and the absorption process are described.
In Sec.~\ref{sec:results} , a numerical analysis is performed to characterize the absorption process.
Finally, in Sec.~\ref{sec:discussion}  the results are discussed and interpreted using analytical approaches.

\begin{figure*}[t!]
    \centering
    \includegraphics[width=15cm]{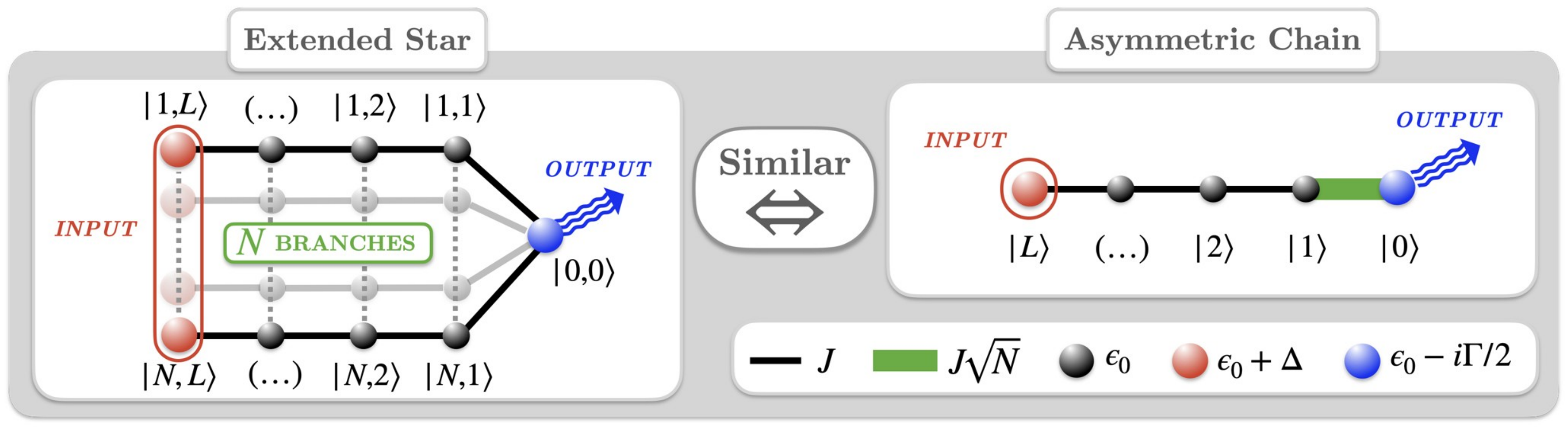}
    \caption{Illustration of the two networks considered in our study. 
    On the left: the extended star which is formed by a central core acting as a trap with an absorbing rate $\Gamma$ (blue site), connected to $N$ branches containing $L$ sites.
    Each branches carries on its extremity an energy defect  with a tunable amplitude $\Delta$ (red sites).
    The exciton can hop between each connected site \textit{via} a hopping constant $J$.
    On the right: the asymmetric chain which carries on its left extremity an energy defect with a tunable amplitude $\Delta$ (red site), and on the other side a trap with absorbing rate $\Gamma$ (blue site).
    The exciton can jump between each connected site \textit{via} a jump constant $J$, except between the two rightmost sites where it is $J\sqrt{N}$.  
    %As demonstrated in Ref.~\cite{yalouz22},  the excitonic dynamics on both networks is equivalent when the exciton start on the input defects (uniform localization on the peripheral sites of the star, and initial exciton localization   on the single defect of the chain).  
    }
    \label{fig:similar_graph}
\end{figure*}

\section{Theoretical Background}\label{sec:theory}

\subsection{ Model Hamiltonians }
 
The two networks we consider are the extended star and the asymmetric chain, as illustrated in Fig.~\ref{fig:similar_graph}. In the absence of the environment, both networks exhibit the same excitonic dynamics provided that the exciton starts from the sites carrying the energy defects.
More precisely, an exciton uniformly distributed over the input sites on the extended star (shown in red) exhibits a time evolution that is isomorphic to that of a single exciton starting from the input site (shown in red too) of the asymmetric chain. 
This similarity has been highlighted in our previous work (see Ref.~\cite{yalouz22}) and will be discussed and illustrated again in the present work (later on in Sec.~\ref{sec:without_deph}).
% Both networks are linked to each other via the following property: on the extended star graph, when considering an exciton uniformly distributed over the input sites (shown in red), the dynamics is isomorphic to that of a single excitation moving on an asymmetric chain. 
In both networks, the resulting dynamics is entirely ruled by a common set of parameters $N,L,J,\Delta$ and $\Gamma$ that define the tight-binding Hamiltonians of each system.
In the following, we will introduce the shape of these Hamiltonians and explain the signification of all the associated parameters.

\subsubsection{Extended star graph}

The excitonic dynamics on the extended star network is described by a standard tight-binding model~\cite{may00} that closely follows the graph's architecture (see Fig.~\ref{fig:similar_graph}).
 Within this model, we assume that each site of the network is occupied by a molecular subunit whose internal dynamics is described by a two-level system. 
 Let $|b,s\rangle$ stand for the state in which the $(b,s)$th two-level system (i.e. site) occupies its first excited state, the other two-level systems remaining in their ground state. 
 Here  $(b,s)$ stand for the $s$th site ($s=1,\ldots,L$) of the $b$th branch ($b=1,\ldots,N$) of the star. 
 The site $(b,s) = (0,0)$ refers to the core of the star. Note that the star involves $\mathcal{N_S} = 1 + NL$ sites.
%  The vacuum state $|\oslash\rangle$ describes all the two-level systems in their ground state.
 Let $\epsilon_0$ denote the excited state energy of the two-level systems except those of the periphery and of the core of the graph.
As illustrated in Fig.~\ref{fig:similar_graph}, the terminal groups are occupied by energy defects whose excitation energy is shifted by an amount $\Delta$ according to $\epsilon_0+\Delta$. 
The core of the graph is occupied by a trap that is responsible for the irreversible decay (i.e. absorption) of the exciton. 
It is characterized by a complex self-energy $\epsilon_0-i\Gamma/2$, where $\Gamma$ defines the exciton decay rate. 
Finally, the exciton can jump between connected nodes \textit{via} a hopping constant $J$. 
 
Following these notations, the Hamiltonian governing the exciton dynamics on the extended star is defined as (the convention $\hbar = 1$ is used throughout the paper)
\begin{equation}
\begin{split}
 {H}  &= \left(\epsilon_0 - i\frac{\Gamma}{2} \right) |00\rangle \langle 00 |   +  \sum_{b=1}^{N}\sum_{s=1}^{L} \left(\epsilon_0+\Delta \delta_{s L} \right) |b, s\rangle \langle b, s |  \\
      &+\sum_{b=1}^{N} J  ( |00\rangle \langle b,1 |+|b, 1\rangle \langle 00 |)  \\
      &+\sum_{b=1}^{N}\sum_{s=1}^{L-1} J  ( |b, s\rangle \langle b, s+1 |+|b ,s+1\rangle \langle b, s |),
\end{split}
        \label{eq:Hstar}  
\end{equation}
where $\delta$ is the Kronecker symbol.

 \subsubsection{Asymmetric chain}
 
 Following the same philosophy, the exciton dynamics on the asymmetric chain (right part of Fig.~\ref{fig:similar_graph}) is described according to a tight-binding model 
\begin{equation} %\label{eq:reduced_hamiltonian}
\begin{split}
 {{H}} &= \sum_{s=0}^{L} \left(\epsilon_0 +\Delta \delta_{s L} -i\frac{\Gamma}{2} \delta_{s0} \right)  \ketbra{s}{s}  \\ 
                   &+ \sum_{s=1}^{L-1} J  ( \ketbra{s}{s+1} + \ketbra{s+1}{s}  )      \\
                   &+ J \sqrt{N}  ( \ketbra{1}{0} + \ketbra{0}{1}  ).
\end{split}
 \label{eq:Hchain}
\end{equation}
Here, the state $\ket{s}$ describes the situation where the exciton occupies the $s$th site of the chain. 
All sites have a local excitation energy $\epsilon_0$ except the two extremities of the chain: the site $s=L$ carries an energy defect $\epsilon_0 + \Delta$, and the site $s=0$ acts as a trap with a complex energy $\epsilon_0-i\Gamma/2$.
All sites are connected to their nearest neighbors with a hopping constant $J$, except the couple of sites $s = 0$ and $1$ connected by an amplitude $J\sqrt{N}$. The presence of the parameter $N$  originates in the equivalence between the dynamics of the asymmetric chain and that of the extended star, as established in details in appendix~\ref{app:connexion}. 
Note that the total number of sites is $\mathcal{N_S} = 1 + L$.

 \subsection{Open Quantum System Dynamics} 

%In the present work, we want to investigate how the presence of an external environment will affect the excitonic quantum transport and the energy absorption. 
To describe the open quantum dynamics of the exciton, we consider a pure dephasing phononic environment modeled by the so-called Haken-Strobl-Reineker master equation \cite{reineker82,haken72,haken73}. 
In this model, the time evolution of the excitonic RDM $\rho(t)$ is governed by the GME 
\begin{equation}
    %  \frac{\partial_t \rho_{xy}}{\partial t} 
     \partial_t \rho_{xy} =  -i \sum_z ({H}_{xz} \rho_{zy} - \rho_{xz} {H}^\dagger_{zy})  -  \gamma (1-\delta_{xy})\rho_{xy},
     \label{eq:QME}
\end{equation}
which is expressed in the local site basis with generic indices $x,y$ and $z$ (a local state $|x\rangle$ refers either to a state $|b,s\rangle$ for the star graph, or to a state $|s\rangle$ for the chain).
Here, $H$ represents the non-hermitian network Hamiltonian, and $\gamma$ is the dephasing rate that is responsible for the irreversible decay of the quantum coherence $\rho_{xy} = \mel{x}{\rho}{y}$ between two distinct sites with generic indices $x$ and $y$.

%\vincent{convention de mukamel $i\dot{\rho}=\mathcal{L} \rho$}
% \saad{Je prefere rester sur la definition de base ou on ne "complexifie" pas la genetatrice. Mukamel c'est un ieuv ! }
 
To simulate the open quantum dynamics of the exciton, one employs the so-called Fock-Liouville space (FLS) method~\cite{mukamel95} which consists in a vectorization of the master equation (see Eq.~(\ref{eq:QME})). 
In this approach, the density matrix  $\rho$  is encoded into a single vector noted $\ket{ \rho \rangle\!}$ (\textit{i.e.}~$\ket{ \rho \rangle\!}$ is a flattened version of the RDM $\rho$).
This allows to rewrite the GME under the form of a generic set of linear differential equations. 
The constant coefficients of this set (associated to the Hamiltonian matrix elements and the dephasing rate) are gathered into the so-called Liouvillian matrix $\mathcal{L}$, and the GME reads then
\begin{equation}
    % \frac{\partial \! \ket{\rho \rangle\!}}{\partial t}
    \partial_t \ket{\rho (t) \rangle\!} = \mathcal{L} \ket{\rho (t) \rangle\!}.
\end{equation}
The formal exponential solution of this equation gives access to the time evolution of the RDM at time $t$ such that
\begin{equation}
    \ket{ \rho(t) \rangle\!} = \exp(\mathcal{L}t) \ket{ \rho(0) \rangle\!},
\end{equation}
where $\exp(\mathcal{L}t)$ is the time-evolution super-operator  
\begin{eqnarray}
\exp(\mathcal{L}t) = \sum_k^{\mathcal{N_S}^2}  \exp(\Lambda_k t) \frac{\ketbra{ \Lambda^{R}_k \rangle\!}{\!\langle \Lambda^{L}_k } }{  \langle\!\langle \Lambda^{L}_k | \Lambda^{R}_k \rangle\!\rangle }, 
\end{eqnarray}
where $\ket{ \Lambda^{L}_k \rangle\!}$ and $\ket{ \Lambda^{R}_k \rangle\!}$ are the left/right eigenvectors of the Liouvillian matrix $\mathcal{L}$, and $  \Lambda_k $ their associated complex eigenvalue. 
Note that the numerical cost of the FLS method scales in $\mathcal{N_S}^6$ which can be problematic when addressing large systems such as the star network.\linebreak
Fortunately, for this network this can be alleviated by using the symmetries of the problem (see Appendix~\ref{app:reduced_set}).

From the knowledge of the time-evolution super-operator, different time-dependent observables may be evaluated. 
In this work, we will mainly focus on the absorption probability $P_A(t)$ expressed as
\begin{equation}
   P_A(t) = 1 -  \text{Tr} \lbrace \rho(t) \rbrace.
   \label{eq:abso}
\end{equation}
The measure $P_A(t)$ describes the probability for the exciton to be absorbed by the trap at time $t$. 
To characterize the absorption process on both networks, we also introduce the absorption time $\tau$ defined as 
\begin{equation}\label{eq:abs_time}
    \tau \longrightarrow P_A(\tau) = 50\%,
\end{equation}
which gives the characteristic time when half of the excitonic population is absorbed by the trap. 
% In practice, the determination of the absorption time $\tau$ is realized via a numerical minimization over $t$ with the \textit{Nealder-Mead} method from the {\textit{scipy}} optimization package.
In practice, the absorption time $\tau$ is determined \textit{via} a numerical minimization (over the parameter $t$) of the cost function  $  \mathcal{C} (t) = (  0.5 -  P_A(t) )^2 $ using the \textit{Nelder-Mead} method from the {\textit{scipy}}  package \cite{scipy}.

%  \saad{
% Note that for the extended system, we can reduced the complexity of the FLS resolution by focusing on some reduced quantities \textit{blabla} SEE APPENDIX VINCENT GENERATRIX
% }

\section{Numerical results}\label{sec:results}

In this  section, the previous formalism is applied to study the absorption process on both the extended star graph and the asymmetric chain. 
In a first part, we will recall the main features observed without dephasing (i.e.~$\gamma=0$). 
%of the excitonic absorption process when no environment is considered and both networks behave the same (i.e. $\gamma=0$). 
In a second part, we will include the environment (i.e. $\gamma>0$) and illustrate the changes occurring on the absorption process.  
Note that in all our simulations, the initial conditions of the dynamics are always the same:  a uniform excitonic delocalization on the star's peripheral defects, and an excitonic localization on the single defect of the asymmetric chain (see Fig.~\ref{fig:similar_graph}).
The hopping constant $J$ is used as the energy reference ({i.e.}~$J = 1$), the absorption rate is fixed to $\Gamma=0.1J$, and we consider $N > 2$.
 
 \subsection{ Absorption without dephasing ($\gamma=0$):\linebreak    optimal energy defects tuning}
 \label{sec:without_deph}

  \begin{figure}[t!]
     \centering\includegraphics[width=7.5cm]{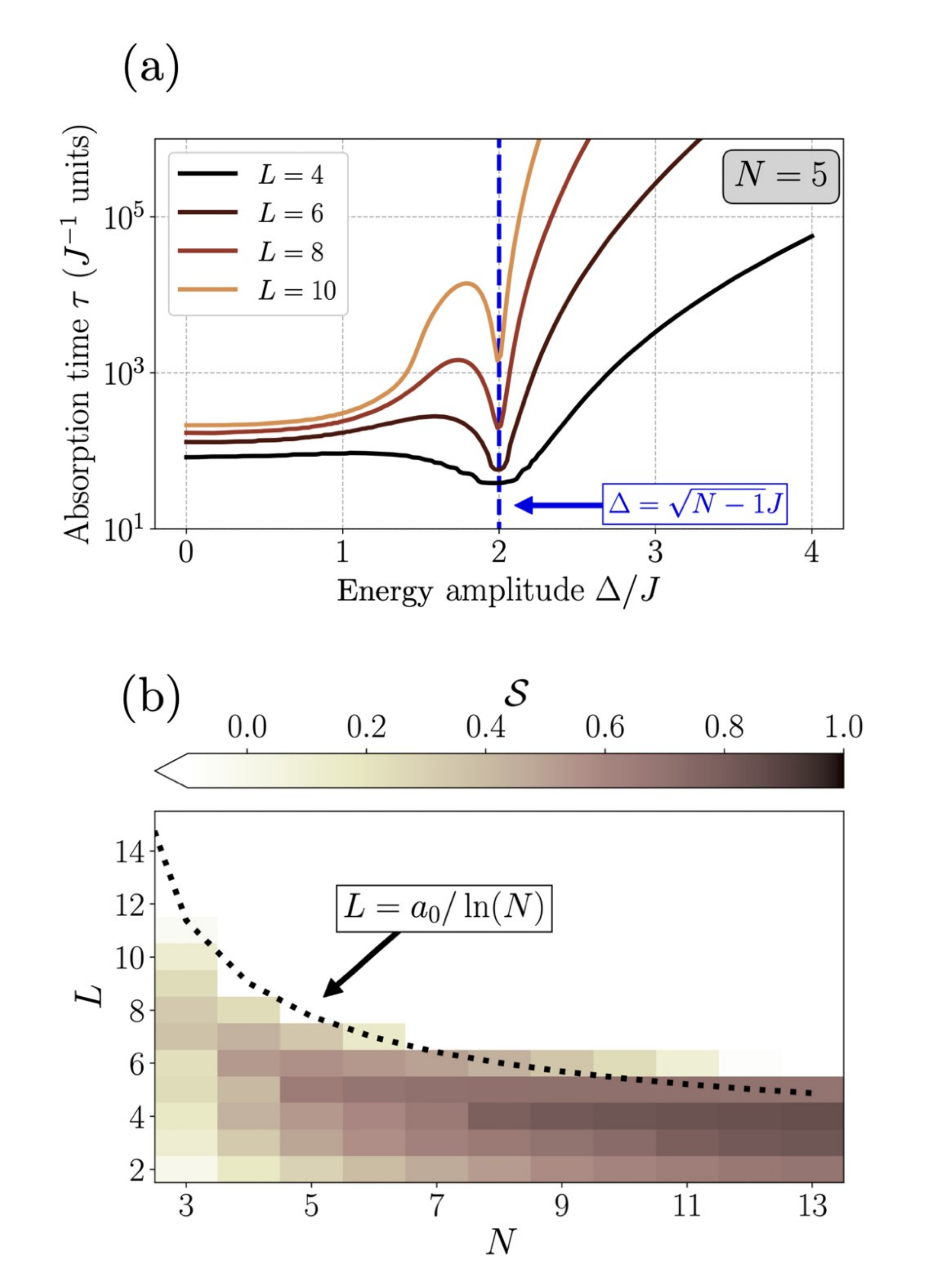}
     \caption{{Evidence of an excitonic absorption speedup by energy defect tuning (with $\gamma =0$).} 
     {(a)} Evolution of $\tau$ as a function of $\Delta$ for $N=5$ and $L=4$, $6$, $8$ and $10$.
     The particular value $\Delta = \sqrt{N-1}J$ where a strong acceleration of the absorption process occurs is marked by a vertical blue dashed line.
     {(b)} Speedup $\mathcal{S}$  (Eq.~(\ref{eq:speedup})) of the absorption process in the $(N , L )$-parameter space.
     Dark colors show where a speedup occurs (\textit{i.e.} $\mathcal{S} > 0$), whereas white area reveal where there is none (\textit{i.e.} $\mathcal{S} \leq 0$).
     The dotted black curve describes the logarithmic limit (see Eq.~(\ref{eq:LB_crit})) defining the region where a speedup is produced.}
     \label{fig:evidence}
 \end{figure}
 
%Let us first recall the main results for the excitonic absorption when no environment is present (\textit{i.e.}  $\gamma=0$).  
When $\gamma=0$, both the extended star and the asymmetric chain behave in the same way. 
In this case, we have shown previously the existence of an optimal value of the energy defect amplitude that is~\cite{yalouz22}
\begin{equation}\label{eq:optimal_conditions}
    \Delta^\text{opt}\approx\sqrt{N-1}J.
\end{equation} 
Around this value, the absorption process presents a strong speedup, namely a reduction of the absorption time $\tau$, provided that the length $L$ is sufficiently short.
To illustrate this feature, we show %in top panel of
in Fig.~\ref{fig:evidence}a the $\Delta$ dependence of the absorption time $\tau$  for  $N=5$ and for $L=4$, $6$, $8$ and $10$. 
As readily seen here, a local minimum of the absorption time systematically appears around the critical value $\Delta^\text{opt}\approx\sqrt{N-1}J$, whatever $L$.
In some cases, this local minimum can go even below the absorption time obtained in the absence of defects (\textit{i.e.} when $\Delta=0$).
This is the case for example with $L \leq 6$  where the absorption time  $\tau$ reaches a global minimum when $\Delta \rightarrow \Delta^\text{opt} $ (see black and brown curves in %top panel of
Fig.~\ref{fig:evidence}a.). 
 However, this is no longer the case for larger $L$ values like $L=8$ or $10$ (see red and orange curves).

 These results highlight the existence of a ``\textit{speedup}'' for the absorption process which depends on both $N$ and $L$. 
To better evidence this feature, we introduce a measure $\mathcal{S}$ of the absorption speedup  that is
 \begin{equation}\label{eq:speedup}
    \mathcal{S} = 1 - \frac{ \tau( \Delta^\text{opt} ) }{ \tau( \Delta=0 ) }.
\end{equation}
With this measure, we evaluate the reduction (or increase) in the absorption time obtained in the presence of energy defects tuned as $\Delta = \Delta^\text{opt}$ compared to the case where no defect is considered (i.e. when $\Delta=0$).\linebreak
By definition, $\mathcal{S} \in ]-\infty,1]$ so that
the closer $\mathcal{S}$ gets to $1$, the more important is the speedup. 
Conversely, $\mathcal{S}<0$ indicates that no speedup is produced at all. 
In the latter case, the absorption time $\tau$ at $\Delta=\Delta^\text{opt}$ is just a local minimum in the $\Delta$-space but not a global one.

%The bottom panel of In 
Fig.~\ref{fig:evidence}b shows a heatmap of $\mathcal{S}$ in the $(L,N)$-parameter space. 
 In this figure, we deliberately choose to rescale the colormap to only highlight regions where a reduction is produced (\textit{i.e.} where $\mathcal{S}>0$) with a gradient of dark color.
 White regions refer to situations in which no reduction in absorption time is produced (when $ \Delta =\Delta^\text{opt} $). 
The present results clearly evidence a region in the $(L,N)$-parameter space where a strong speedup $S>0$ is produced. 
This region can be delimited by the following rule for the structural parameter $L$ and $N$: 
 \begin{equation}\label{eq:LB_crit}
     \mathcal{S}>0 \Longrightarrow L \leq L^*  \text{, with } L^* \approx \frac{a_0}{\ln(N)},
 \end{equation} 
where $a_0 = 12.5$ is a numerical coefficient. For both networks we see here that the architectural parameter $L$ is the main limiting factor for the appearance of the absorption process speedup. 
% The logarithmic nature of the law given in Eq.~(\ref{eq:LB_crit}) is in fact linked to the presence of specific excitonic eigenstates that tend to localize exponentially on both the defects and the absorbing site, generating an effective energy transfer between the two regions.  
% This point has been extensively addressed in our previous work~\cite{yalouz22} and, for the sake of brevity, we refer the interested reader to that paper for a more detailed discussion.
%This point has been extensively addressed an explained in our previous work~\cite{yalouz22}. 
%For the sake of brevity in this reminder section, we refer the interested reader to this work for a more detailed discussion on this point.

%As a conclusion for this first section, we will summarize the main properties observed in the absence of dephasing. For both networks, there exist optimal tuning conditions for the energy defects which allows to optimize the energy absorption process. If we set the energy defects like  $\Delta^\text{opt} = \sqrt{N-1}J$, the absorption time $\tau$ can be strongly reduced compared to when $\Delta=0$. This optimization law only works for networks presenting a structure parameter $L \leq L^*$ (with $L^*$ given in Eq.~(\ref{eq:LB_crit})). 

\subsection{Absorption with dephasing ($\gamma> 0$)} 
\label{sec:with_deph}

In this second section, we include the presence of a dephasing environment ($\gamma> 0$) and we illustrate the changes involved in the absorption process.

\begin{figure}[h!]
    \centering 
    \includegraphics[width=7.5cm]{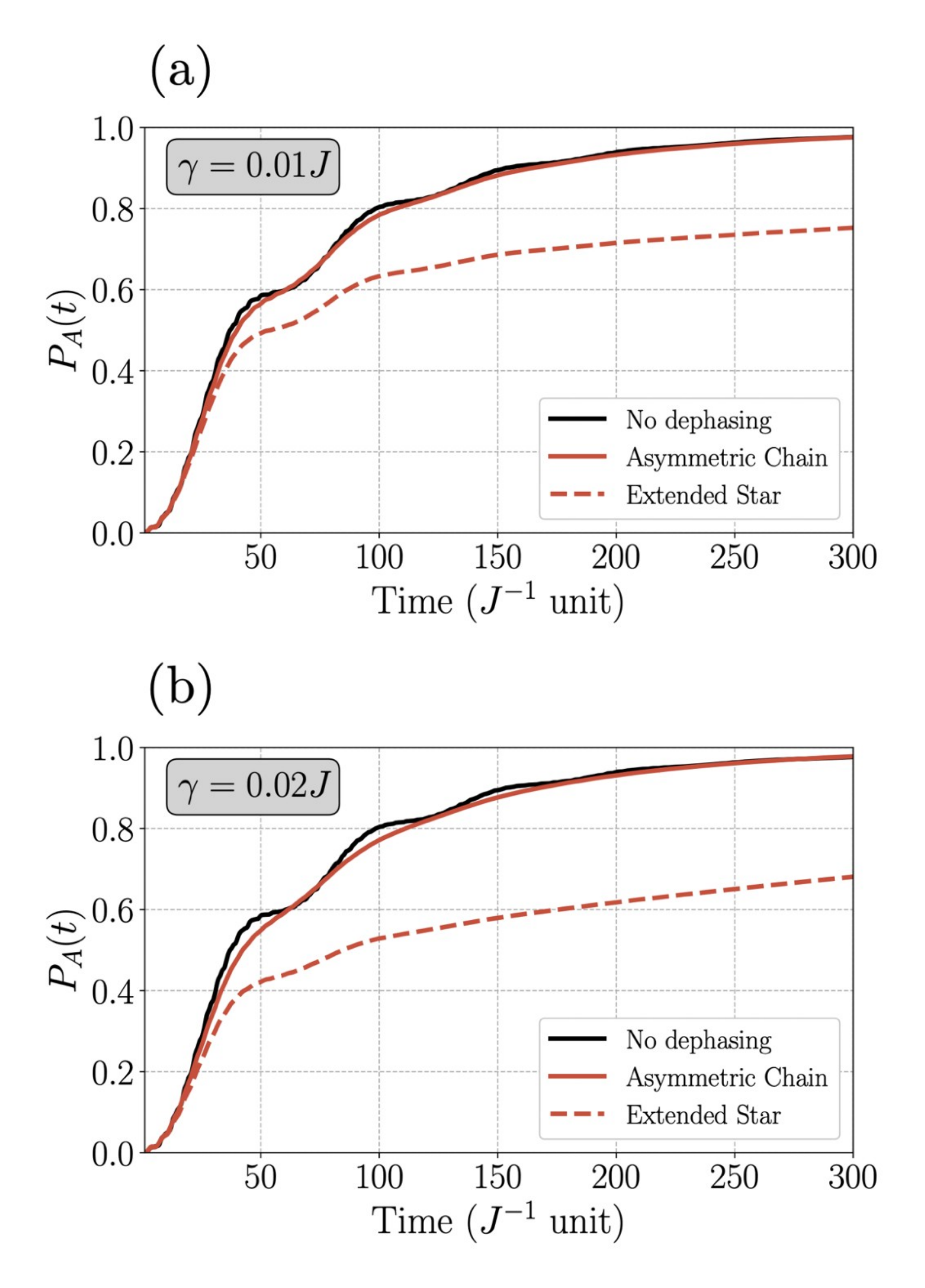}
    \caption{{Time evolution of the absorption probability $P_A(t)$  in both networks with dephasing $\gamma >0$ ($N=L=4$ and $\Delta^\text{opt}=\sqrt{N-1}J$).} 
    {(a)}  weak dephasing $\gamma  = 0.02J$.
    {(b)}  moderate dephasing $\gamma  = 0.1J$.
    Red full lines are results for the asymmetric chain while dashed curves are for the extended star.
    Black full lines show the reference result when no dephasing is considered and both networks admit a same absorption process.
    }
    \label{fig:Comparison}
\end{figure}

\subsubsection{Evidence of different time-scales for the absorption processes on the two networks}

Let us first highlight an important feature observed in all our simulations: in the presence of the dephasing, the time-scale for the realization of the absorption process is very different between both networks.\linebreak
 To illustrate this feature, we show in Fig.~\ref{fig:Comparison} the time evolution of the absorbed population  $P_A(t)$  for both networks with weak ($\gamma = 0.02J$ in Fig.~\ref{fig:Comparison}a ) and moderate ($\gamma = 0.1J$ in Fig.~\ref{fig:Comparison}b) dephasing. 
 In this figure, dashed red lines illustrate the results obtained for the extended star, while full red lines are used for the asymmetric chain.
 As a reference, we also show with full black curves  $P_A(t)$  when $\gamma= 0$ (i.e. the case where both networks behave the same).   
 
In Fig.~\ref{fig:Comparison}a, we see that the presence of even a very weak dephasing amplitude %(here $\gamma = 0.02J$) 
leads to a different time evolution of the absorption process, depending on the network considered.
For the asymmetric chain,  $P_A(t)$ is barely affected by the dephasing, and the resulting absorption time $ 45 J^{-1}$ is still quite close to that obtained without dephasing $ 37 J^{-1}$.
%(see the solid red curve compared with the black curve).
Conversely, more significant changes occur in the case of the extended star.
Here, the absorption process is slowed down and the absorption time becomes $\tau \sim 75 J^{-1} $, %(dashed red curve)
which is almost twice as long as in the case where no dephasing is considered.
As shown in Fig.~\ref{fig:Comparison}b, this slow-down  becomes even more important when  the dephasing rate reaches $\gamma = 0.1J$.
In this case, the absorption time on the extended star increases considerably to $\tau \sim 200 J^{-1}$ (four times more than in the reference case without dephasing). 
Conversely, the asymmetric chain shows only a very moderate increase in absorption time, which becomes $50 J^{-1}$ (still fairly close to the case without dephasing).

\subsubsection{Evolution of the energy absorption optimization law in the $\Delta$-parameter space}
\label{sec:Delta_space}

\begin{figure}[t!]
    \centering
    \includegraphics[width=7.5cm]{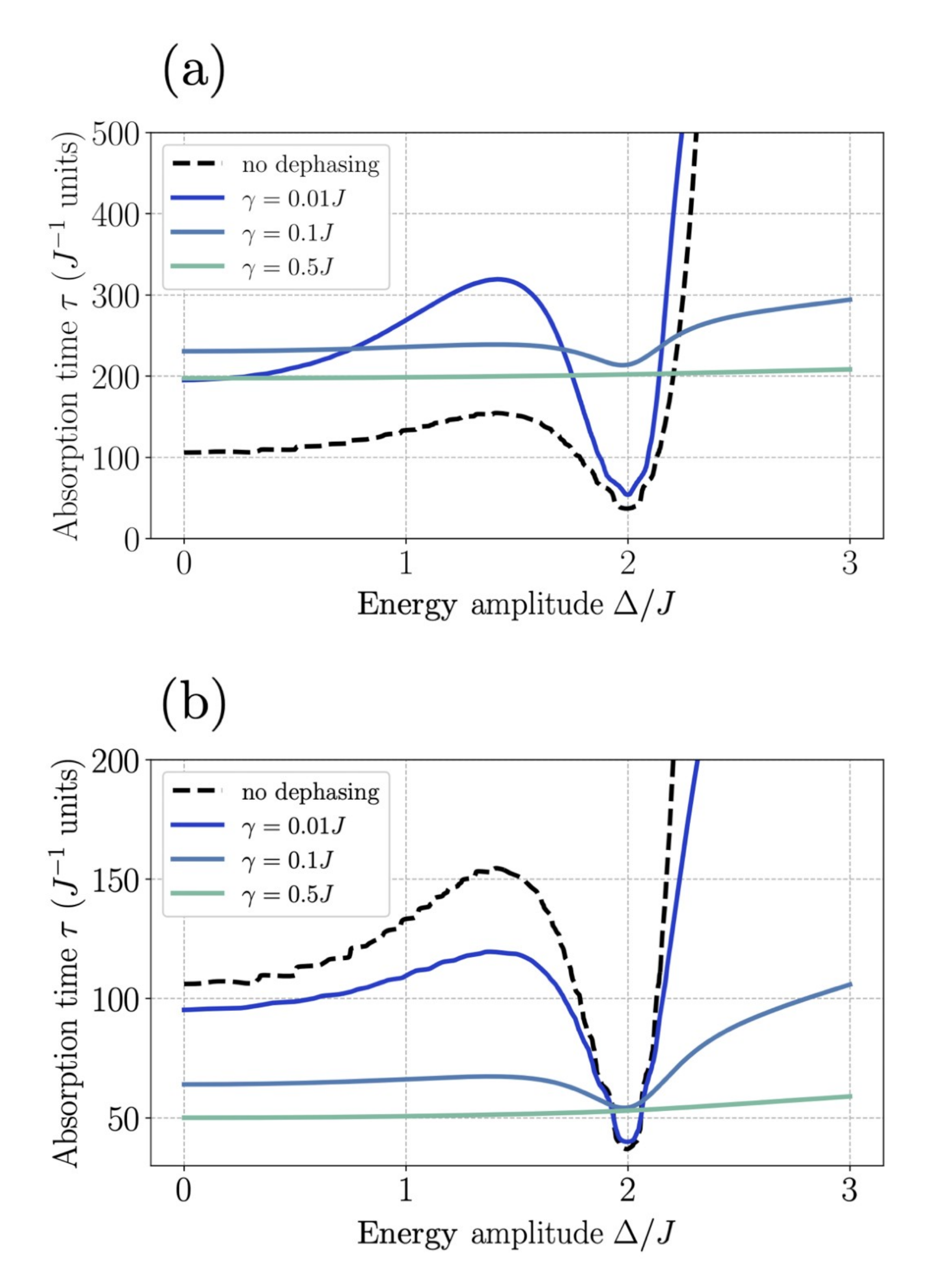}
    \caption{{Evolution of the absorption time $\tau$ with $\Delta$ for increasing values of dephasing rate $\gamma =0.01J, 0.1J  $ and $0.5J$} (with $N=L= 5$). {(a)}~Results for the extended star. {(b)}~Results for the asymmetric chain.}
    \label{fig:Delta_evo_with_deph}
\end{figure}

At this point, the fundamental question arises:
is the energy-defect tuning $\Delta= \sqrt{N-1}J$ still optimal for the energy absorption in the presence of the environment ? 
To address this question, the $\Delta$-dependence of the absorption time in different dephasing regimes is shown in Fig.~\ref{fig:Delta_evo_with_deph} for the extended star (Fig.~\ref{fig:Delta_evo_with_deph}a) and for the chain (Fig.~\ref{fig:Delta_evo_with_deph}b). We consider $N=L= 5$ and a color gradient is used to illustrate the behaviors obtained for three increasing dephasing amplitudes $ \gamma = 0.01J$, $0.1J$ and $0.5J$.
As a reference, we also show here the absorption time obtained when $\gamma=0$ with a black dashed curve.  

By comparing the results obtained for both networks, we see that the two systems present similarities and differences.
% If we focus on top panel, one can readily see that the local minimum of the absorption time arising for $\Delta = \sqrt{N-1} J$ is affected by the presence of the dephasing following three phases. 
First of all, let us highlight the most important property shared by the two systems: the survival of the optimal energy defect amplitude $\Delta = \sqrt{N-1}J$. 
Indeed, for both networks the $\Delta$-dependent absorption time curves still exhibit a global minimum for this specific  energy defect amplitude, as long as the dephasing amplitude is $\gamma \leq 0.1J$ (i.e.~weak regime). 
Beyond this value (moderate/strong dephasing regime), the minimum disappears and no optimization is produced.
Secondly, we can see from Fig.~\ref{fig:Delta_evo_with_deph} that increasing $\gamma$ tends to converge the profile of the curves around different central values.
%Indeed, when  $\gamma=0.01J \rightarrow 0.5J$ the overall amplitude of the $\Delta$-dependant curves tend to compress and converge towards $\sim 200J^{-1}$ for the extended star, and $\sim 50J^{-1}$ for the asymmetric chain.
In the strong dephasing limit, the absorption time becomes almost $\Delta$ independent. It converges towards $\sim 200J^{-1}$ for the extended star, and $\sim 50J^{-1}$ for the asymmetric chain.
Note that these two limits values compare well with the order of magnitude defined by the analytical ratio $\ln(2)\mathcal{N_S}/\Gamma$ which gives here $180 J^{-1}$ for the extended star and  $42 J^{-1}$ for the chain.
We will show in Sec.~\ref{sec:discussion} that this analytical ratio is a key quantity for interpreting the evolution of the absorption time with the presence of dephasing.
Finally, let us mention that after converging  around different limit values, an increase of the dephasing amplitude like $\gamma \gg 0.5J$ (very strong regime) tends to generate an overall shift of the absorption time curves towards greater values for both networks (not shown here).
Note that the $\Delta$ dependence of the absorption time $\tau$ is very general and is not limited to the case presented in Fig.~\ref{fig:Delta_evo_with_deph}. 
Indeed, many other numerical simulations (not shown here) have revealed that all the features presented here  (i.e. survival of the optimization law, flattening of the curves, etc.)  are recurrent features arising for both systems.

 \subsubsection{Assessing the survival of the excitonic absorption optimization law in the $(L,N)$-parameter space}

\begin{figure}[t!]
    \centering 
    \includegraphics[width=\columnwidth]{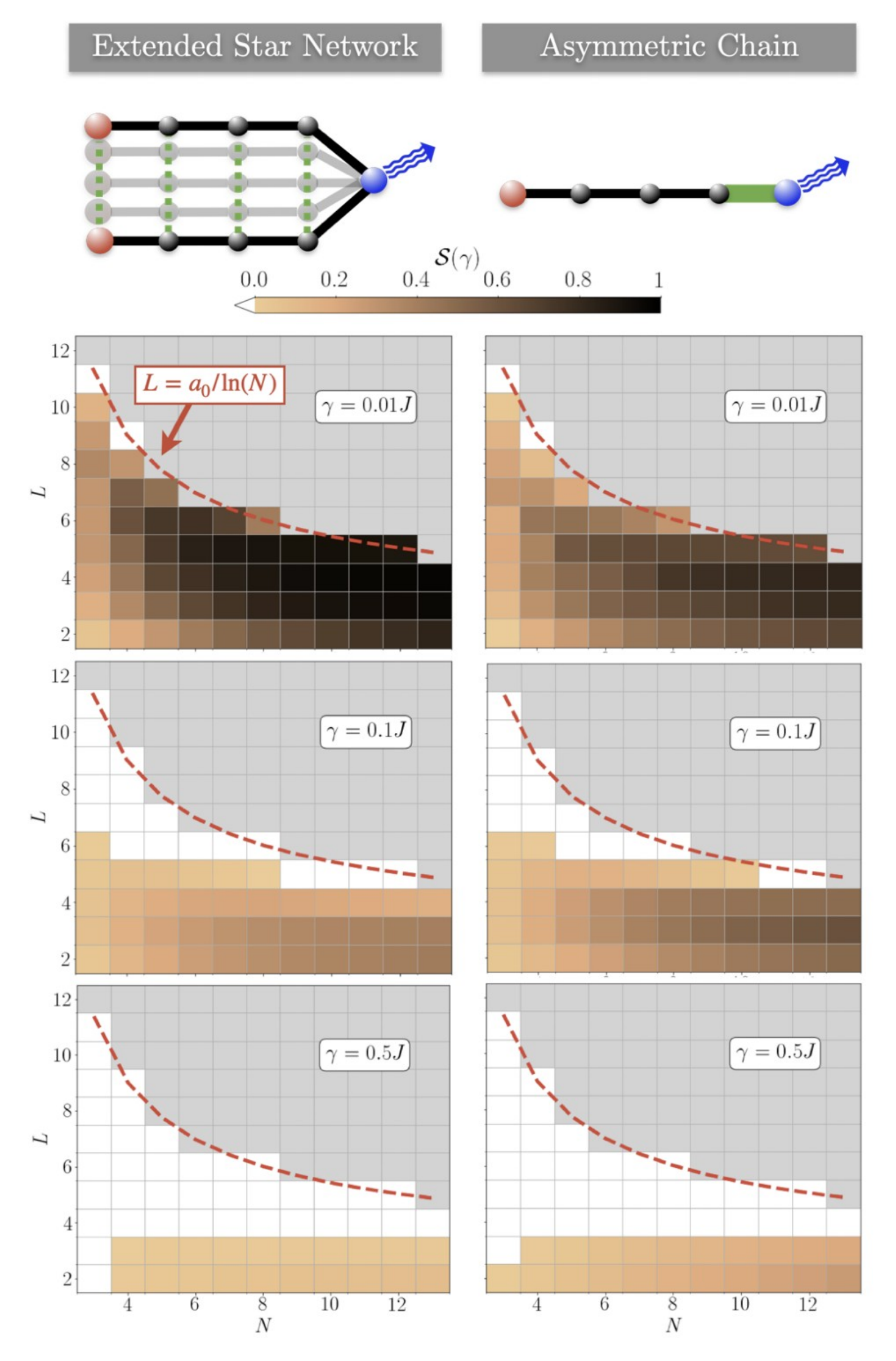}  
    \caption{Evolution of the dephasing-dependent measure $S(\gamma)$ (as given in Eq.~(\ref{eq:gamma_speedup})) in the  $(L,N)$-parameter space for both networks and increasing values of dephasing rate. Left/right column respectively shows the results for the extended star and the asymmetric chain. Top to bottom rows show results for $\gamma =0.01J$, $0.1J  $ and $0.5J$. 
    A color gradient (ranging from orange to black) is used to show the regions where the absorption optimization mediated by the energy defects is still present (i.e. $S(\gamma) > 0 $).
    % Here, $\epsilon$ is a very small threshold amplitude fixed to $\epsilon=10^{-3}$.
    White color is used everywhere  the optimization is lost $S(\gamma) \leq 0$.
    Note that we only focus here on the region below the red dashed curve where the optimization was originally detected in the absence of dephasing (see Sec.~\ref{sec:without_deph}).
    The rest of the space is thus represented in grey. } 
    \label{fig:heatmaps}
\end{figure}

All our results suggest that both networks present some similarities regarding the survival of the optimal setting of energy defects in the presence of dephasing.
\linebreak
% In both cases, the absorption time  $\tau$ exhibit a local minimum  when $\Delta=\sqrt{N-1}J$, as long as $\gamma$ remains small enough (typically $\gamma <0.1J)$.
% The only main difference between both networks lies in the presence of a global shift for $\tau$ that is more pronounced in the case of the extended star than in the asymmetric chain.
In this final part of our numerical study, we  show that these similarities actually hold for different $(L,N)$ configurations. 
To demonstrate this point, we introduce a $\gamma$-dependent measure of the absorption time speedup $\mathcal{S}(\gamma)$ that reads
 \begin{equation} 
    \mathcal{S}(\gamma) = 1 - \frac{ \tau(\gamma , \Delta^\text{opt}) }{ \tau(\gamma , \Delta=0 ) }.
    \label{eq:gamma_speedup}
\end{equation}
This measure is inspired by the one introduced in Eq.~(\ref{eq:speedup}) (with $\gamma=0$) with the difference that it now includes the effect of the dephasing on the absorption time.
We use this measure as an indicator for assessing the survival of the optimal tuning $\Delta^\text{opt}=\sqrt{N-1}J$ of the energy defects.
Here, $\mathcal{S}(\gamma) > 0$ indicates a survival of the optimal tuning, i.e. the absorption time still presents a minimum when $\Delta=\Delta^\text{opt}$.
Conversely, $\mathcal{S}(\gamma) \leq 0$ indicates that the minimum has been suppressed  and that no optimization can be produced. 
% Here, $\epsilon = 10^{-3}$ is a small threshold amplitude used to detect when the difference in the absorption time with/without energy defects is negligible.
% Here,  $\mathcal{S}(\gamma)$ provides an estimate (for a fixed value of dephasing $\gamma >0$) of the reduction (or augmentation) of the absorption time obtained in the presence of defects tuned like $\Delta = \Delta^\text{opt}$ compared to the case when no defects is considered $\Delta=0$.
% This measure lives in the domain $\mathcal{S}(\gamma) \in ]-\infty,1 ]$.
Note that $\mathcal{S}(\gamma)$ only evaluates the survival of the minimum for the absorption time, but it doesn't quantify the presence of a global offset as discussed in section~\ref{sec:Delta_space}.
 In Fig.~\ref{fig:heatmaps}, we illustrate the evolution of $\mathcal{S}(\gamma)$ in the $(L,N)$-parameter space for increasing values of dephasing (from $\gamma=0.01J$ to $0.5J$) with a series of heatmaps. 
Left/right columns respectively show the results obtained for the extended star network and the asymmetric chain.
Top to bottom rows show results from weak to moderate dephasing.
Note that one only focuses on the $(L,N)$-parameter subspace where an optimization was initially detected in the absence of dephasing (below the red dashed curve defined by Eq.~(\ref{eq:LB_crit})). 
The rest of the $(L,N)$-parameter space is thus colored in grey.  
The orange-black gradient shows regions where optimization is still present, while the white areas show regions where it has been lost.

As readily seen in Fig.~\ref{fig:heatmaps}, both networks exhibit very similar behaviors for the evolution of $\mathcal{S}(\gamma)$ as $\gamma$ increases. 
For weak dephasing $\gamma =0.01J$ (see first row in Fig.~\ref{fig:heatmaps}),  we observe a positive amplitude $\mathcal{S}(\gamma) > 0$ almost everywhere in the $(L,N)$-parameter space which indicates the survival of the optimization for all the configurations.
Then, when the value of dephasing increases to $\gamma \sim 0.1J$ (see second row in Fig.~\ref{fig:heatmaps}), one can see that this general property doesn't hold anymore. 
Around this value of dephasing, one starts to detect the arising of $(L,N)$ configurations for which the measure becomes null/negative $\mathcal{S}(\gamma) \leq 0$, as indicated by the arising of large white areas below the red curve. 
When the dephasing reaches the moderate regime with $\gamma =0.5J$,  we observe an (almost) complete transition in the $(L,N)$-parameter space (third row in Fig.~\ref{fig:heatmaps}).
In the latter case, the white zone extends over a large part of the $(L,N)$ configurations, revealing that the absorption optimization mediated by energy defects at $\Delta = \sqrt{N-1}J$ is lost on both networks.\linebreak
Note that additional simulations (not shown here) have shown that the absorption optimization completely disappears when one increases the dephasing to even larger values $\gamma \gg 0.5J$ (i.e. the white area spreads everywhere).

\begin{figure*}[t!]
    \centering
    \includegraphics[width=17cm]{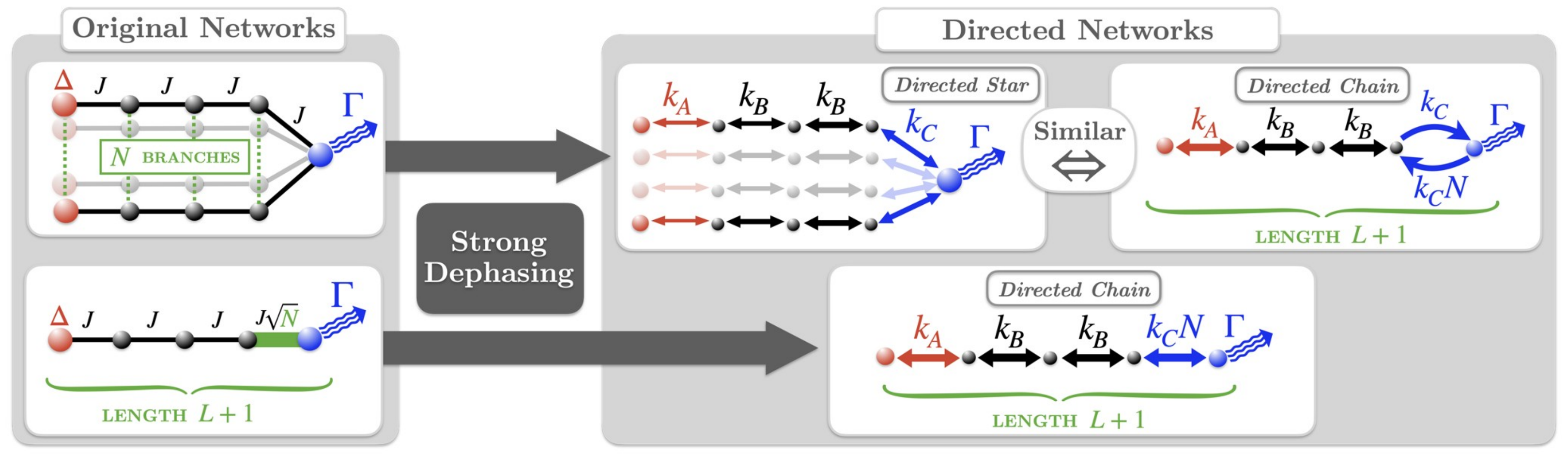}
    \caption{Illustration of the change of behavior for the two networks considered in our study in the presence of strong dephasing. 
    When $\gamma$ is assumed to be large enough,  both networks act as directed chains with rates $k_A, k_B, k_C$ and $\Gamma$. 
    The two directed networks are very similar, except for the rates connecting the trap to its first neighbour (right part of the chains).  }
    \label{fig:directed_net}
\end{figure*}

\section{ Discussion }\label{sec:discussion}

Our numerical study allowed to evidence a series of features.
Firstly, it has been shown that the presence of dephasing rescales the time evolution of the energy absorption process depending on the network under consideration. 
The process becomes globally much slower for the extended star than for the asymmetric chain.
% This is linked to the appearance of a global offset affecting the $\Delta$-dependent absorption time  $\tau$ (i.e. a delay), which is greater in the case of the extended star than for the asymmetric chain.
Secondly,
% in spite of this difference,
one observed that both   networks  present very similar behaviors regarding the survival of the energy absorption optimization law. 
In both cases,  a minimum of the absorption time $\tau$ still arises around the value of energy defect $\Delta^\text{opt}=\sqrt{N-1}J$, as long as the dephasing remains weak enough. 
% Note that this minimum is globally higher than the one observed in the absence of dephasing ($\gamma=0$) due to the presence of a global offset that affects the absorption time $\tau$ when $\gamma >0$.
% This offset is globally larger for the extended star than for the asymmetric.
% In this case, the main difference between the two networks consists in the arising of a global shift of the absorption time curve (more important in the extended star than in the asymmetric chain).
Finally, when $\gamma$ increases to reach intermediate/strong dephasing regimes (typically $\gamma > 0.1J $), it has been shown  that the optimization law doesn't hold anymore in both networks  whatever the $(L,N)$-configuration considered.

\subsection{Excitonic diffusion view:\linebreak from undirected to directed networks}
\label{sec:diffusive}

To interpret our numerical results, we realize an analytical development based on the strong dephasing limit. 
% When the dephasing becomes sufficiently large, the excitonic dynamics loses its coherent character, leading to so-called \textit{exciton diffusion}.
In this context, the GME Eq.~(\ref{eq:QME})  can actually be replaced by a classical ``\textit{rate equation}'' that reads
\begin{equation}\label{eq:rate_eq}
    \partial_t \bold{P}(t) =    \bold{K} \bold{P}(t).
\end{equation}
This equation describes the time evolution of the excitonic populations $\rho_{ss}$ on the sites of a network, which are collected into the vector $\bold{P}$ of size $\mathcal{N_S}$ (with $P_s = \rho_{ss}$).
Here, $   \bold{K}$ is the matrix whose elements encode the different rates governing the kinetic of the excitonic populations.
As explained in Ref.~\cite{silbey09}, the determination of the excitonic rates in the strong dephasing limit is achieved following several steps.
First, one starts from the GME expressed in the site basis (see Eq.~(\ref{eq:QME})). 
Then, assuming that the dephasing is strong enough, we consider stationary conditions for the  quantum coherences  ($\partial_t \rho_{xy} = 0$) between nearest neighbor sites originally connected by a hopping constant (see Fig.~\ref{fig:similar_graph}).
Every other ``long range'' quantum coherences are neglected. 
Next, the short-range coherences $ \rho_{xy}$ are expressed in terms of the populations $\rho_{xx}$ and $\rho_{yy}$. 
Using this redefinition, one can finally rewrite the GME under the form of a purely classical rate equation (as given by Eq~(\ref{eq:rate_eq})) that governs the kinetic of the excitonic site populations.

By applying this scheme to the extended star and the asymmetric chain, one shows that the two networks transform into two pretty similar \textit{``directed chains''} as illustrated in Fig.~\ref{fig:directed_net}.  
These directed chains include three types of rates noted $k_A, k_B$ and $k_C$ (plus the irreversible trapping $\Gamma$) that read
\begin{equation}
    k_A = \frac{2J^2 \gamma}{\gamma^2 + \Delta^2}, \quad k_B =\frac{2J^2}{\gamma}, \quad  k_C =\frac{2J^2  }{\gamma+\Gamma/2  }.
    \label{eq:def_rates}
\end{equation}
% The latter include the effect of the different parameters of the systems.
Note that the form of these rates (obtained from the scheme introduced in Ref.~\cite{silbey09}) are similar to the ones present in other works focusing on excitonic diffusion {(see for example Refs.~\cite{zhang17,zhangdisorder17,Leegwater96} and references therein)}. 
As illustrated in Fig.~\ref{fig:directed_net}, the transformation of the asymmetric chain into a directed chain is straightforward.
However, this is not the case for the extended star for which two steps are required.
First, the extended star is transformed into a \textit{directed star} (as shown on the right part of Fig.~\ref{fig:directed_net}) following the steps described previously. 
Then, starting from the associated rate equation, one introduces a change of variable to replace the total population of each set of symmetrical sites of the directed star (following the circular symmetry of the network) by a single effective population.
The resulting $L+1$ effective populations will be associated to the effective sites of the \textit{directed chain} illustrated in top right of Fig.~\ref{fig:directed_net}. % illustrated in Fig.~\ref{fig:directed_net}. 
% To do this, we use a change of variable so that each effective site of the directed chain encode the total population of a set of symmetrical sites on the directed star (following the circular symmetry of the network). 
% In Fig.~\ref{fig:directed_net}, we illustrate the structure of the resulting directed chain.
% The meaning of each effective sites on this chain is then the following:  the leftmost site encodes the total population of the directed star's peripheral defects, then the second site encodes the total population of the defects' first neighbors (etc.), up to  the rightmost site of the directed chain which encodes the population of the directed star's core.  
Using this mapping, one can show that the resulting directed chain exhibits a very special feature: an anisotropy in the rates linking its two rightmost effective sites (see in  Fig.~\ref{fig:directed_net}).  
The excitonic transfer towards the trapping core is governed by a rate $k_C$ which is $N$ times lower than the rate $Nk_C$ governing the transfer in the opposite direction.
Comparatively, note that the directed network obtained for the asymmetric chain doesn't present such an asymmetry (here both rates are $k_CN$ as shown in Fig.~\ref{fig:directed_net}).
This exotic phenomenon of anisotropic rate has already been highlighted in several works on dendritic architecture (see Refs.~\cite{bar-haim97,Monthus82}), and is in fact related to the connectivity of the central site, which is greater than that of the branch sites on the directed star.
% For more details about these properties, we refer the interested reader to Appendix \ref{app:star_network} where we provide an illustrative example of how a directed star network (with $L=1$) maps onto an effective directed chain which exhibits anisotropic rates.

%  \begin{figure*}[t!]
%     \centering
%     \includegraphics[width=15cm]{Final_discussion.pdf}
%     \caption{{Evolution of the absorption time $\tau$ with the dephasing rate $\gamma$ for different values of $(L,N)$ parameters (with $\Delta= \sqrt{N-1} J$).
%     \textbf{Left column:} extended star network. \textbf{Right column:} asymmetric chain.
%     The results on the top row are for a configuration $(N=6,L=3)$, whereas the bottom row shows results for $(N=3, L=6)$.}}
%     \label{fig:Abso_time_analytical}
% \end{figure*}

\begin{figure*}[t!]
    \centering
    \includegraphics[width=15cm]{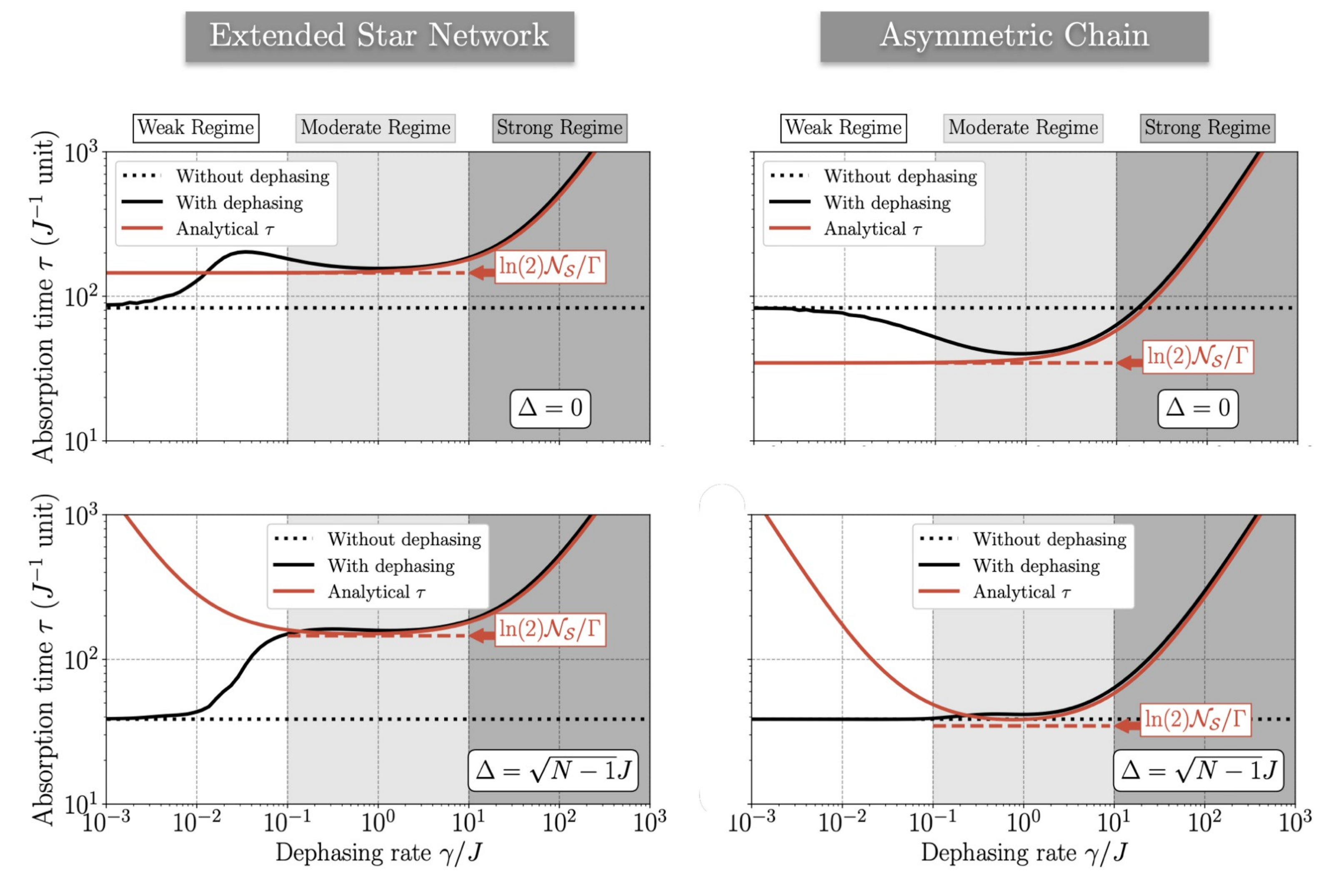}
    \caption{{Evolution of the absorption time $\tau$ with the dephasing rate $\gamma$ for different values of energy defects amplitudes $\Delta$ parameters (with $N=5,L=4$).
    {Left column:} extended star network. {Right column:} asymmetric chain.
    The results on the top row are for a configuration $\Delta = 0$, whereas the bottom row shows results for $\Delta = \sqrt{N-1}J$.}}
    \label{fig:Abso_time_analytical}
\end{figure*}

\subsection{Evolution of the absorption process in the presence of dephasing}

Starting from this mapping on directed chains, one can then derive an analytical form for the absorption time.
This is indeed possible in the case of directed chains using approaches based either on the analytical properties of the inverted rate matrix $ \bold{K}^{-1}$, or on waiting time distribution formalism  (see Refs.~\cite{bar-haim98,silbey08,silbey09,gopich03} and references therein).
We refer the interested readers to Appendices~\ref{app:MFPT} and~\ref{app:WTDF} where we provide more details on these methods. 
Based on this, one can obtain an analytical form of the excitonic absorption time for both directed chains that reads $ \tau \approx \ln(2)  \Bar{\tau}$,  with 
\begin{equation}
     \Bar{\tau}   =  \frac{\cal{N_S}}{\Gamma} +\frac{1}{k_A}   +  \frac{(L+1)(L-2)}{2k_B} +  \frac{L}{  k_C}     \left(1 
 + \frac{1-N}{N}  \delta^{Ch} \right).   %+ \frac{L}{N k_C } .
     \label{eq:MFPT_analytical}
\end{equation}
% \begin{equation}
%      \Bar{\tau}   =  \frac{\cal{N_S}}{\Gamma} +\frac{1}{k_A}   +  \frac{(L+1)(L-2)}{2k_B} + \frac{L}{k_C}    %+ \frac{L}{N k_C } . 
% \end{equation}
% \saad{J'ai corrigé en effet c'est bien $1/k_A$ et $L/k_C$ !!! Et il faut encore que je readapte car il existe une petite diff au niveau de $k_C$}
Here, $k_A, k_B$ and $k_C$ are the rates defined in Eq.~(\ref{eq:def_rates}), and $\delta^{Ch}$ is a term equal to one only for the asymmetric chain (zero otherwise).
% In the analytical form of the absorption time given in Eq.~(\ref{eq:MFPT_analytical}), two differences can be observed in between both networks studied.
% The first one  is related to the first term $ \mathcal{N_S}/\Gamma$ in Eq.~(\ref{eq:def_rates}) which depends on the total size $\cal{N_S}$ of a given network ($\mathcal{N_S}=1+L$ for the asymmetric chain, and $\mathcal{N_S} = 1+NL$ for the extended star).
%  We will show in the following that this contribution plays a key role in the interpretation of all the results observed in our numerical simulations presented in Sec.~\ref{sec:with_deph} when increasing the dephasing amplitude. 
% The second difference in Eq.~(\ref{eq:MFPT_analytical}) is linked to the very last term.
% Numerical simulations (not shown here) have revealed that this difference of contribution is not important to understand what is occurring when we increase the dephasing amplitude to the inter mediated regime (it is only important in the very strong dephasing regime)
Note that the first term $ \mathcal{N_S}/\Gamma$ depends on the total size $\cal{N_S}$ of the given network ($\mathcal{N_S}=1+L$ for the asymmetric chain, and $\mathcal{N_S} = 1+NL$ for the extended star).
 % We will show in the following that this contribution is the key to understand the behaviors observed in our numerical study (see Sec.~\ref{sec:with_deph}). 
% Note also in Eq.~(\ref{eq:MFPT_analytical}) a difference in the last term which yields $L/k_C$ for the asymmetric chain, and $L/(Nk_C)$ for the extended star.
% Numerical simulations (not shown here) have revealed that this difference only plays a role in the very strong dephasing regime, and is therefore not important in interpreting the changes in behavior highlighted in our numerical study.
In the following, we will use the analytical form Eq.~\ref{eq:MFPT_analytical} to interpret the absorption time evolution with the dephasing. 
To this end, we illustrate  in Fig.~\ref{fig:Abso_time_analytical} the numerical/analytical results of the variations of $\tau$ as a function of $\gamma$.
White, grey and dark grey areas are used to respectively delimit weak, moderate and strong dephasing regimes. 
 Solid red lines show the analytical results  Eq.~(\ref{eq:MFPT_analytical}), while solid black curves are for pure numerical simulation.
Left and right columns respectively show results for the extended star and the asymmetric chain.
% at the optimal value $\Delta=\sqrt{N-1}$. 
% Top/bottom rows show the result for two different $(L,N)$ configurations with $(L=6,N=3)$ on the top and  $(L=3,N=6)$ shown at the bottom.
Top/bottom rows show results for two energy defects amplitudes with $\Delta = 0$ (top), and $\Delta = \sqrt{N-1}J$ (bottom).
As a reference, we also report on this plot the absorption time obtained  without dephasing (horizontal black dashed lines).
% Note that this unperturbed time is the same on each row of Fig.~\ref{fig:Abso_time_analytical} which facilitates the comparison between both networks in the presence of dephasing (see the same black dashed horizontal line). 

% We will now detail the scenario of evolution of the absorption time when the dephasing increases.
% As readily seen in Fig.~\ref{fig:Abso_time_analytical}, the analytical results given by Eq.~(\ref{eq:MFPT_analytical}) fit very well the numerical simulation when $\gamma \geq 0.1 J$ (i.e. the intermediate/strong dephasing regime represented by the grey area) whatever the networks and the value $\Delta$ considered.
% This correspondence will help us to interpret the scenario of evolution of the absorption time as we increase the dephasing.
% Based on this observation, one can then better interpret how the absorption process evolve when the dephasing amplitude $\gamma$ increases.  
To begin, let us focus on the weak dephasing regime $\gamma < 0.1J$ (white area in Fig.~\ref{fig:Abso_time_analytical}).
% In this case, the excitonic transport is still strongly coherent and no big change occurs: the value of $\tau$ remains very close to the absorption time obtained in the absence of dephasing $\gamma=0$ (see the black dashed line).  
As readily seen in the plot, the analytical results fail here to reproduce the numerical simulations.  
This discrepancy is in fact expected as in this regime the dephasing is not dominant: the excitonic transport is still strongly coherent (Sec.~\ref{sec:diffusive}).
With the survival of coherent excitonic transport, it is therefore natural to expect the survival of the optimal energy transfer. 
This goes along the lines of what has been observed in our numerical study (see Sec.~\ref{sec:with_deph}) regarding the survival of the absorption optimization law in both networks for $\gamma <0.1J$. 
Interestingly, even if the excitonic dynamics is still fundamentally coherent in this regime, the presence of dephasing will nevertheless affect the time-scale of the absorption process. 
Indeed, we see in Fig.~\ref{fig:Abso_time_analytical} that an increase in $\gamma$ causes the absorption time $\tau$ (solid black curve) to deviate progressively from the reference time obtained without dephasing (black dashed line). 
Here, the amplitude of $\tau$ gradually converges towards a plateau that is reached typically when the dephasing enters the moderate dephasing regime $ \gamma \rightarrow  0.1J$ (see horizontal red dashed lines in all panels).

The regime of moderate dephasing $\gamma \in [0.1J, 10J]$ (central grey region in Fig.~\ref{fig:Abso_time_analytical}) is precisely where a turning point occurs in the nature of excitonic dynamics.
In this case, the dephasing amplitude $\gamma$ becomes comparable to the hopping constant $J$, so that the diffusive transport point of view begins to be valid (see Sec.~\ref{sec:diffusive}).
This can be clearly observed in Fig.~\ref{fig:Abso_time_analytical}, as here the analytical results start to match very well with the numerical simulations. 
This holds  whatever the network and the value $\Delta$ considered.
Interestingly, we see in this regime that the absorption time systematically converges towards a plateau value. 
The latter depends on the network, but it remains the same whatever $\Delta$ (see rows of a same column in Fig.~\ref{fig:Abso_time_analytical}).
This behaviour is consistent with what was highlighted earlier in the $\Delta$-parameter space in Sec.~\ref{sec:Delta_space} (see Fig.~\ref{fig:Delta_evo_with_deph}).   
Based on the good correspondence between the analytical/numerical behaviors here, one can actually relate the amplitude of this plateau to the first term of Eq.~(\ref{eq:MFPT_analytical}), that is $\tau = \ln(2)\mathcal{N_S}/\Gamma$  (which represents an approximate low bound of the equation).  \linebreak
With this analytical form, one can better understand the features evidenced in our numerical study when $\gamma \sim J$. 
Firstly, the $\mathcal{N_S}$-dependence of the plateau explains why we observe a different time-scale in the absorption process occurring on the two networks when the dephasing increases.
Indeed, the total number of sites for the extended star $\mathcal{N_S}=1+LN$ is larger than for the asymmetric chain  $\mathcal{N_S}=1+L$. 
The resulting plateau is thus higher in the first case than in the second. 
As a consequence, the closer we get to the moderate dephasing regime, the slower will be the absorption process for the star compared to the asymmetric chain.
This goes along the lines of what has been observed in our numerical study (see Fig.~\ref{fig:Comparison} and Fig.~\ref{fig:Delta_evo_with_deph}).
Secondly, it is interesting to note that the plateau value $\tau = \ln(2)\mathcal{N_S}/\Gamma$ is $\Delta$ independent. 
As a result, we understand here that the absorption time will always converge in the moderate dephasing regime to the same plateau whatever $\Delta$ (as shown in Fig.~\ref{fig:Abso_time_analytical}).
This property explains why previously in Fig.~\ref{fig:Delta_evo_with_deph} (Sec.~\ref{sec:Delta_space}) we observed a global flattening of the absorption time curves in the $\Delta$-parameters space when $\gamma \rightarrow 0.5J$.
% \saad{petite interpretation ici}
% To conclude and to complete the discussion, let us note that the analytical form of the plateau value $\tau = \ln(2)\mathcal{N_S}/\Gamma$ occurring in the moderate dephasing regime may in fact be accessible \textit{via} other methods than the diffusion view introduced in Sec.~\ref{sec:diffusive}. 
% We explain in appendix~\ref{app:demonstration_limit_value} how we also managed to recover this amplitude using perturbation theory in Fock-Liouville space, based on knowledge of the steady state of the purely dephasing HSR model.

Finally, when   the strong dephasing regime is reached (dark grey area in Fig.~\ref{fig:Abso_time_analytical}), a second turning point occurs for the excitonic dynamics.
In this case, the dephasing is so strong that the excitonic transfer gets hindered, leading to the slow-down of the energy absorption process.
This can be readily seen in Fig.~\ref{fig:Abso_time_analytical} with the increase of the absorption time $\tau$ with the dephasing amplitude $\gamma$.\linebreak
Here again, note the excellent correspondence between the analytical results and the numerical simulations.
Based on this, one can understand that the behavior of the absorption time is actually described by the three additional contributions of Eq.~(\ref{eq:MFPT_analytical}) that depend on the rates $k_A$, $k_B$ and $k_C$ (as defined in Eq.~(\ref{eq:def_rates})). \linebreak
Globally, these contributions reveal that the absorption time $\tau$ scales (approximately) linearly with the dephasing amplitude, i.e. $\tau \propto \gamma$ when $\gamma \gg J$.  
Note the presence of a difference in the last term of Eq.~(\ref{eq:MFPT_analytical}) which yields $L/k_C$ for the star, and  $L/(Nk_C)$ for the asymmetric chain.
This reveals that, in the case of the
asymmetric chain, the absorption time increases less rapidly with $\gamma$ than for the star.
This feature can directly be related to the  anisotropic rates present in the directed chain associated to the extended star (see Fig.~\ref{fig:directed_net}).
In this case, the excitonic energy transfer is biased: the transfer with the rate $k_C$  towards the trap is $N$ times weaker than in the opposite direction where the rate is $Nk_C$. 
To conclude our discussion, one should mention that the arising of the general exciton transfer slow-down in the strong dephasing regime may be in fact linked to the arising of the so-called ``quantum Zeno effect''~\cite{Misra77} (also known as the ``watch-dog effect'').
Here, the associated interpretation is the following: a strong dephasing amplitude $\gamma$ forces the localization of the exciton on its initial position, preventing then its efficient transfer to the trap.

\section{ Conclusion  }

In this work, we studied the open quantum system dynamics of an exciton and its absorption on two networks: an extended star and an asymmetric chain. 
The specificity of these two networks was recently highlighted in  Ref.~\cite{yalouz22} where we show that both architectures actually present a similar energy absorption process that can be optimized by the inclusion of tunable energy defects (in the absence of environment).
As a direct extension to this work, our investigations here focused on the question of how this absorption optimization is affected by the presence of dephasing, and also if in practice the two networks  respond differently to the presence of an environment.

Our work revealed that both networks actually present very similar behaviors concerning the survival of the optimization law when we increase the dephasing. 
In both cases, the  absorption process can still be optimized by the use of energy defects as long as the dephasing remains weak enough.  
In this regime, the only difference between the two networks is the appearance of an offset in the absorption time, which is more significant in the case of the extended star than in the asymmetric chain.
Analytical developments based on the diffusive limit of the excitonic transport allowed to connect this behavior to the size of the networks (i.e. the number of sites $\mathcal{N_S}$).     
Additionally, our results have shown that for both networks the optimization law is totally lost as soon as we enter moderate/strong dephasing regimes.
In this case, analytical developments revealed that the absorption time increases linearly with the dephasing amplitude. 
This feature can be interpreted as the arising of the so-called quantum Zeno effect.

Therefore, our work evidenced the possibility for both networks to exhibit an absorption optimization mediated by energy defects in the presence of an environment (to some limit). 
These results naturally motivate new questions that could represent interesting starting points for future developments. 
For example, it would be interesting to see if the similarities shared by the two networks would resist to the presence of a static disorder (e.g. site energies and/or hopping constants modulations).
In this context, the symmetry breaking induced by disorder should open new paths on the networks for the excitonic dynamics leading to potential different absorption mechanisms.
On a different note, it would be also interesting to see if more realistic excitonic environment modeling would lead to new behaviors (e.g. full quantum description of the phonons).
Indeed, in the current work we employed a Linblad master equation (i.e. HSR model) which is fundamentally markovian and thus excludes the possibility to observe any long time quantum coherence in the system.
This description prevents us from observing potential exotic non-markovian effects that would naturally arise in the excitonic transport.
All these ideas will be considered in future projects and papers.

%    Cao, Jianshu, and Robert J. Silbey.\textbf{ "Optimization of exciton trapping in energy transfer processes."} The Journal of Physical Chemistry A 113.50 (2009): 13825-13838.

%  Leegwater, Jan A. \textbf{"Coherent versus incoherent energy transfer and trapping in photosynthetic antenna complexes."} The Journal of Physical Chemistry 100.34 (1996): 14403-14409.

%  Pearlstein, Robert M. \textbf{"Exciton migration and trapping in photosynthesis."} Photochemistry and Photobiology 35.6 (1982): 835-844.
 
%  Zhang, Yang, et al. \textbf{"Opening-assisted coherent transport in the semiclassical regime."} Physical Review E 95.2 (2017): 022122.

%  Random walks on the Bethe lattice. BD Hughes, M Sahimi - Journal of Statistical Physics, 1982

% Random walk on the Bethe lattice and hyperbolic Brownian motion. C Monthus, C Texier - Journal of Physics A: Mathematical and ..., 1996 - iopscience.iop.org 

\appendix

\section{Similarity in the excitonic transport on the extended star network and the asymmetric chain}
\label{app:connexion}

Let us consider the extended star graph with $N$ branches of length $L$ on which the exciton dynamics is governed by the Hamiltonian $H$ Eq.~(\ref{eq:Hstar}).
$H$ being invariant under the discrete rotation of angle $\theta= 2 \pi /N$ and centered on the core of the star, its diagonalization is simplified when one works with an intermediate basis involving the state localized on the core $|0,0\rangle$  and a set of orthogonal Bloch states $|\chi^{(k)}_s \rangle $ with $s=1,2,...,L$ and $k = 1,2,...,N$. A given Bloch state is defined as
\begin{equation}
|\chi^{(k)}_s \rangle=\frac{1}{\sqrt{N}} \sum_{b=1}^{N} e^{-ik b \theta} |b,s \rangle.
\end{equation}
Within this basis, $H$ is expressed as a direct sum $H=H^{(1)} \oplus H^{(2)}...\oplus H^{(N)}$  where $H^{(k)}$ is the block Hamiltonian associated to the good quantum number $k$. For all $k \neq N$, all the block $H^{(k)}$ are identical. They are expressed as 
\begin{equation}\label{eq:k_diff_NB}
\begin{split}
H^{(k \neq N)} &= \sum_{s=1}^{L} (\epsilon_0+\Delta \delta_{s L}) |\chi^{(k)}_s \rangle \langle \chi^{(k)}_s | \\
                       &+  \sum_{s=1}^{L-1} J  ( |\chi^{(k)}_{s+1} \rangle \langle \chi^{(k)}_s |+|\chi^{(k)}_{s} \rangle \langle \chi^{(k)}_{s+1}|). 
\end{split}
\end{equation}
For $k=N$, the Hamiltonian $H^{(N)}$ is defined as 
\begin{equation} \label{eq:k_eq_NB}
\begin{split}
    H^{(N)} &= \left(\epsilon_0-i\frac{\Gamma}{2} \right) |0,0\rangle \langle 0,0| \\
              &+ \sum_{s=1}^{L} (\epsilon_0+\Delta \delta_{s L}) |\chi^{(N)}_s \rangle \langle \chi^{(N)}_s | \\
              &+ \sqrt{N} J (|0,0\rangle \langle \chi_1^{(N)} |+|\chi_1^{(N)} \rangle \langle 00| ) \\
              &+ \sum_{s=1}^{L-1} J ( |\chi^{(N)}_{s+1} \rangle \langle \chi^{(N)}_s |+|\chi^{(N)}_{s} \rangle \langle \chi^{(N)}_{s+1}|).
\end{split}
\end{equation}
One sees here that when the exciton is initially uniformly delocalized over the peripheral sites of the star (i.e. with an initial state $|\chi^{(N)}_{s} \rangle$), its dynamics is confined in the $k=N$ subspace. Restricting our attention to that subspace, the notations can be simplified 
by renaming the basis vectors as
\begin{equation}
  |s) = \left\{
      \begin{aligned}
        |0,0\rangle   &     \ \ \textrm{if} & s=0 \\
        |\chi_s^{(N)}\rangle &    \ \ \textrm{if} & s>0. \\
      \end{aligned}
    \right.
\end{equation}
Within these notations, the restriction of the Hamiltonian in the $k=N$ subspace is rewritten as
\begin{equation} \label{eq:reduced_hamiltonian}
\begin{split}
 H^{(N)} &= \sum_{s=0}^{L} \left(\epsilon_0-i\frac{\Gamma}{2} \delta_{s0}+\Delta \delta_{s L} \right)  | s) (s|  \\ 
                   &+ \sum_{s=1}^{L-1} J  (|s)(s+1| + |s+1)(s|) \\
                   &+ J \sqrt{N}  ( |0)(1| + |1)(0|).
\end{split}
\end{equation}
Eq.~(\ref{eq:reduced_hamiltonian}) corresponds to the Hamiltonian given in Eq.~(\ref{eq:Hchain}) which governs the exciton dynamics on the asymmetric chain shown in Fig.~\ref{fig:similar_graph}.

\section{ Reduced set of equations for the excitonic dynamics on the extended star network  }
\label{app:reduced_set}

Studying the absorption process on the extended star needs the knowledge of the time evolution of the absorption  probability $P_A(t)$ Eq.(\ref{eq:abso}). Because, it depends only on the trace of the exciton RDM $\rho(t)$, its characterization does not require to resolve the full set of $\mathcal{N_S}^2$ coupled equations describing the dynamics of the GME Eq.(\ref{eq:QME}) (which numerically scales in  $\mathcal{N_S}^6$ for the FLS method). 
Instead, one can only focus on a reduced number of quantities thus reducing drastically the numerical cost needs to solve the problem.

% To start, let us recall the Liouville-von Neumannequation governing the excitonic dynamics in the site basis 
% \begin{equation}
%     \partial_t \rho_{xy} =  -i \sum_z (\mathcal{H}_{xz} \rho_{zy} - \rho_{xz} \mathcal{H}^\dagger_{zy})  -  \gamma (1-\delta_{xy})\rho_{xy}
% \end{equation}
% % \begin{equation}
% %     \partial_t \rho =  i(\mathcal{H} \rho - \rho \mathcal{H}^\dagger)  -   \mathcal{R}\rho
% % \end{equation}
% where $\mathcal{H}$ is the effectif non-hermitian hamiltonian 
% \begin{equation}
%     \mathcal{H} = H - i \frac{\Gamma}{2} \ketbra{00}{00}
% \end{equation}
% that encodes the irreversible excitonic trapping effect at the core of the graph with a rate $\Gamma$ (see Eq.~(\ref{eq:Hstar})). 
% In the local site basis, the equation reads
% \begin{equation}
%     \partial_t \rho_{xy} =  -i \sum_z (\mathcal{H}_{xz} \rho_{zy} - \rho_{xz} \mathcal{H}^\dagger_{zy})  -  \gamma (1-\delta_{xy})\rho_{xy}
% \en d{equation}
To proceed, let us define the four key quantities we should focus on. 
First, the core population
\begin{eqnarray}
P_0 = \rho_{00,00}.
\end{eqnarray}
Second, the peripheral intra-branches populations and coherences
\begin{eqnarray}
P_{ss'} = \sum_{b=l}^{N} \rho_{bs,bs},
\end{eqnarray}
which represent a $(L\times L)$ matrix whose elements are independent on the number of branches and depend only on the size of the branches. 
The diagonal elements contain the populations of each generation of the graph while the non-diagonal elements characterize the coherences between sites of the same branch.  
Third, the inter-branches coherences
\begin{eqnarray}
C_{ss'} = \sum_{b=1}^{N}\sum_{b'=1}^{N} (1-\delta_{bb'})\rho_{bs,b's'} ,
\end{eqnarray}
which represent a $(L\times L)$ matrix whose elements characterize the coherences between different branches only. 
And finally, the core-branches coherences 
\begin{eqnarray}
K_{s} = \sum_{b=1}^{N}  \rho_{bs,00} ,
\end{eqnarray}
which represents a column vector of dimension $L$ whose elements characterize the coherences between the core and the branch sites.

Using these key quantities, the excitonic dynamics is governed by a reduced set of $n_S=1+ 2L+ 2L^2$ coupled equations reading
\begin{widetext}
\begin{equation}
\begin{split}
\partial_t P_0     &= -\Gamma P_0 -iJ (K_1 - K_1^*)\\
\partial_t P_{ss'} &= -i \sum_{s"} (h_{ss"}P_{s"s'}  - P_{ss"}h_{s"s'}) + iJ (K_s\delta_{s',1} - K_{s'}^*\delta_{s,1}) - \gamma P_{ss'}(1-\delta_{ss'}) \\
\partial_t C_{ss'} &= -i \sum_{s"} (h_{ss"}C_{s"s'}  - C_{ss"}h_{s"s'}) + iJ (N-1)(K_s\delta_{s',1} - K_{s'}^*\delta_{s,1}) - \gamma C_{ss'} \\
\partial_t K_s     &= -\left(\frac{\Gamma}{2} + \gamma \right) K_s -i \sum_{s"} h_{ss"} K_{s"} + iJ (P_{s1} + C_{s1}) -i J NP_0 \delta_{s1}\\
\partial_t K_s^*     &= -\left(\frac{\Gamma}{2} + \gamma \right) K_s^* + i \sum_{s"} h_{ss"} K_{s"}^* - iJ (P_{1s} + C_{1s}) +i J NP_0 \delta_{s1},\\   
\end{split}
\label{eq:reduced_set}
\end{equation}
\end{widetext}
where $\bold{h}$ represents a $(L \times L)$ reduced  hamiltonian matrix defined as:
\begin{equation}
    \bold{h} = 
    \begin{pmatrix}
    0 & J &  &   &  \\
    J & 0  & J    &   &  \\
     & J  &  \ldots   &   &  \\
     &    &      & 0 & J  \\
     &    &      & J & \Delta  \\
    \end{pmatrix}.
\end{equation}

The reduced set of differential equations  Eq.~(\ref{eq:reduced_set}) can be put into a compact matrix form
\begin{equation}
    \partial_t \bold{v} = \bold{G} \bold{v},
\end{equation}
where $\bold{v}$ is a column vector (dimension $n_S$) containing all the elements of the key quantities introduced before (the matrices have been flattened), and $\bold{G}$ is the square generatrix matrix encoding the coefficients of the differential equations (with dimension $ n_S^2$).
A left/right eigen decomposition can be conducted on $\bold{G}$ to exponentiate the associated matrix and consequently to solve the time evolution of the key quantities as 
\begin{equation}
    \bold{v}(t) = \exp({\bold{G} t}) \  \bold{v}(0).
\end{equation}
With this approach, the absorbed population at time $t$ is obtained by summing the pertinent elements of the vector $\bold{v}$ returning
\begin{eqnarray}
P_A(t) = 1 - P_{0}(t) - \sum_{s} P_{ss} (t).
\end{eqnarray}
Note that the numerical cost of the FLS method based on this reduced set of equations will  scale in  $n_S^3$, which is drastically lower compared to the original cost in $\mathcal{N_S}^6$ (i.e. exact diagonalization of a matrix with dimension $n_S$ instead of $\mathcal{N_S}^2$).

\section{Analytical form of the excitonic absorption time on a directed linear network using the properties of the inverse rate matrix}
\label{app:MFPT}

\begin{figure}[h!]
    \centering
    \includegraphics[width=8cm]{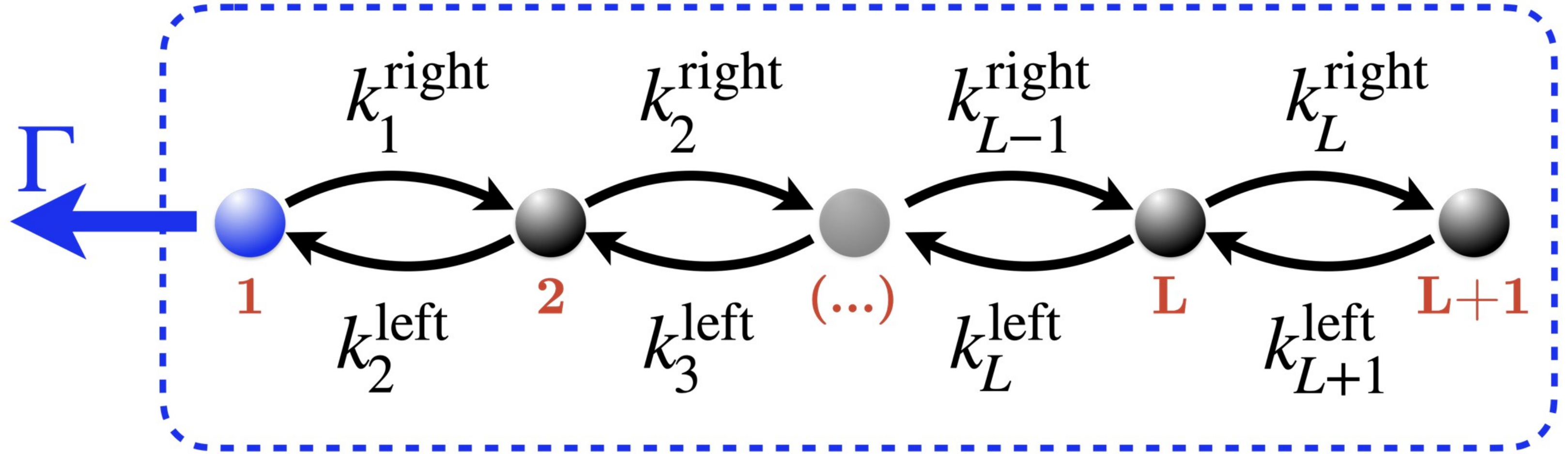}
    \caption{Directed chain of length $\mathcal{N_S} = L+1$ with an irreversible trapping process occurring on the first site with a rate $\Gamma$.}
    \label{fig:chain_dir}
\end{figure}

 When considering an intermediate/strong dephasing ($\gamma \geq J$),  the excitonic dynamics on both networks reduces to a random walk on a directed chain with a trap, as depicted in~Fig.~\ref{fig:chain_dir}.
 Starting from this view,  we will detail in this appendix how to estimate the excitonic absorption time using the concept of ``\textit{Mean First Passage Time}'' (MFPT)~\cite{Polizzi16}.

The MFPT, noted $\bar{\tau}$, defines the typical time for a random walker starting from an initial node of a given directed graph  to reach another specific distant targeted node.
 If the target node contains a trap, $\bar{\tau}$ can be seen as the typical time required for the random walker to initiate its irreversible trapping process.
 In our study, we will focus on the case of a random walker evolving on the directed chain of length $L+1$ as shown in Fig.~\ref{fig:chain_dir}. 
 We will assume here that the initial node occupied by the walker is on the right extremity of the chain (site $L+1$). 
 The targeted node will be the one at the other extremity of the chain (site 1) where an irreversible trapping process takes place (with rate~$\Gamma$).  
 On this chain, the walker hops between nodes with left/right oriented rate constants noted $k_s^\text{left}$/ $k_s^\text{right}$ with ``$s$'' the index of the site from which the hops come from.
 All this information is collected in a rate matrix $ \bold{K}$ that reads 
  \begin{equation} \label{eq:shape_K}
     \bold{K} =  
     \begin{bmatrix} 
         -\Gamma - k^\text{right}_1 & k^\text{left}_2  &    &      &   \\
         k^\text{right}_1          & - k^\text{right}_2- k^\text{left}_2 & k^\text{left}_3 &    &  \\
                                    & k^\text{right}_2 & \ldots &   \ldots &   \\
                                   &  &  \ldots &   \ldots & k^\text{left}_{L+1}\\
                                   &  &   &      k^\text{right}_{L}& - k^\text{left}_{L+1}\\
     \end{bmatrix}.  
 \end{equation}
From the knowledge of $   \bold{K}$, the time evolution of the random walker is formally given by 
\begin{equation}
    \bold{P} (t) = \exp(   \bold{K}t) \bold{P}(0),
\end{equation}
where $\bold{P}(0) = (P_1(0),P_2(0),\ldots,P_{L+1}(0))^T$ is a column vector that encodes the initial sites populations (walker starting on site $L+1$), and $\bold{P} (t)$ the resulting time-evolved vector at time~$t$.
Within these notations, the MFPT is defined as

 \begin{equation}
     \Bar{\tau} =  \sum_\text{ s=1 }^\text{L+1}  \int_0^{+\infty}  P_\text{s}(t) dt. 
     \label{eq:integral}
 \end{equation}
Formally, this quantity  can be seen as the sum of the residence time of the walker on all the sites (see Ref.~\cite{Polizzi16}), which provides an estimate of how long the walker ``survives'' on the network before starting to feel the effect of the trap on site $1$.
Due to the irreversible trapping, we know that for $t \rightarrow +\infty $ we have $\exp(Kt) \rightarrow 0$.
Using this property, one can integrate Eq.~(\ref{eq:integral}) and shows that the MFPT takes the following form
\begin{equation}
    \Bar{ \tau }    =  - \sum_{s=1}^{\text{L+1}} [   \bold{K}^{-1} \bold{P}(0)]_s.
\end{equation}
As shown here, the elements of the inverse rate matrix $   \bold{K}^{-1}$ provide an estimate of the MFPT.
In practice, getting access to the analytical form of $   \bold{K}^{-1}$ is a very difficult task.
However, it has been shown (see Ref.~\cite{bar-haim98}) that this information can be exactly accessed in the case of directed chains \textit{via} recurrence relations.
In this case, the exact form of $ \bold{K}^{-1}$ allows us to derive an analytical form of the MFPT that reads (see Eq.~(19) in Ref.~\cite{bar-haim98})
\begin{equation}
    \Bar{\tau}   =  \sum_{m=1}^{L+1} r_m +\sum_{m=1}^{L}  \frac{1}{r_m k^\text{right}_m} \sum_{n=m+1}^{L+1}  r_n ,
     \label{eq:MFPT_chain}
\end{equation}
with the amplitudes $r_m$ defined as
\begin{equation}
    r_m   =  
    \begin{cases}
         \dfrac{1}{\Gamma}, &\text{for } m = 1 .\\
          \dfrac{1}{\Gamma} \dfrac{ k_1^\text{right}k_2^\text{right}\ldots k_{m-1}^\text{right} }{  k_2^\text{left}k_3^\text{left}\ldots k_{m}^\text{left} },  &\text{for } m > 1.
    \end{cases}  
\end{equation}
We can determine the shape of these amplitudes for the two  directed chains associated to the extended star network and the asymmetric chain. 
In the case of the extended star we have
\begin{equation}
    r_m   =  
    \begin{cases}
        1/\Gamma, &\text{for } m = 1 .\\
        N/\Gamma ,  &\text{for } m > 1.
    \end{cases} 
\end{equation}
While for the case of the asymmetric chain we have
\begin{equation}
    r_m   =   \dfrac{1}{\Gamma}, \text{for } m \geq 1.
\end{equation}
By injecting these amplitudes in Eq.~(\ref{eq:MFPT_chain}), one can finally derive the analytical form of the MFPT $\Bar{\tau}$ given in Eq.~(\ref{eq:MFPT_analytical}) for both networks.

% \section{Temperature for a dephasing rate}

% A way to relate a value for dephasing to a temperature found in the ENAQT paper : 

% One can estimate the dephasing rate as a function of temperature by employing a standard system–reservoir model [42]. In this context, the spectral density is given by 
% \begin{eqnarray}
% J(\omega) = \sum_i \omega_i^2 \lambda_i^2 \delta (\omega-\omega_i)
% \end{eqnarray},
% where $\omega_i$ are frequencies of the harmonic-oscillator bath modes and $\lambda_i$ are dimensionless couplings to the respective modes.
% In the continuum limit, we assume an Ohmic spectral density with cutoff such as : 
% \begin{eqnarray}
% J(\omega) = E_R \hbar \omega_C \omega \exp(-\omega/\omega_C).
% \end{eqnarray}
% For the FMO complex,
% the reorganization energy is found to be ER = 35 cm-1 and the cutoff $\omega_C = 150$ cm-1. 
% In the Markovian regime, the dephasing rate $\gamma$ is given as the
% zero-frequency limit of the Fourier transform of the bath correlator. As a result, $\gamma$ is found
% to be proportional to the temperature and the derivative of the spectral density at vanishing
% frequency,
% \begin{eqnarray}
% \gamma_{deph} = 2\pi \frac{kT}{\hbar} \frac{\partial J(\omega)}{\partial \omega}|_{\omega=0} 
% \end{eqnarray} 
% Using the the previously defined spectral density, this relation yields :
% \begin{eqnarray}
% \gamma_{deph} = 2\pi \frac{kT}{\hbar}  \frac{E_R}{\omega_C}
% \end{eqnarray} 
% This gives a rough estimate for the dephasing rate at room
% temperature of about 300 cm-1 in the case of the FMO complex.

\section{ Analytical form of the excitonic absorption time on directed linear networks using the waiting time distribution formalism}
\label{app:WTDF}

%%%%%%%%% INTRO A REFAIRE 

In complement to appendix~\ref{app:MFPT}, let us  describe an alternative approach to determine the absorption time in the strong dephasing limit. 
This approach is based on the ``\textit{waiting time distribution}'' formalism (see Ref~\cite{silbey08}), which is a mathematical tool that has been used in various fields ranging from excitonic transport~\cite{silbey09} to single-molecule chemical chain reactions~\cite{silbey08,gopich06,kolomeisky00,singh2022,kou05}.

\begin{figure}[h!]
    \centering
    \includegraphics[width=8cm]{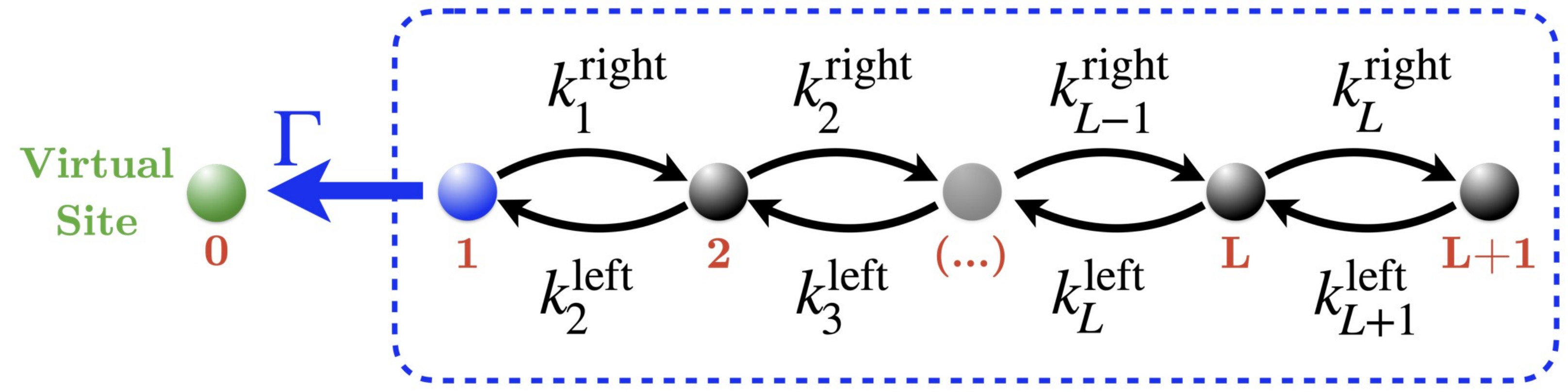}
    \caption{Directed chain of length $\mathcal{N_S} = L+2$ with an irreversible trapping process occurring between sites $1 \rightarrow 0$ with a rate $\Gamma$.}
    \label{fig:WTDF}
\end{figure}

Within this approach the exciton kinetic is described by a linear directed chain of length $L+2$ as illustrated in Fig.~\ref{fig:WTDF}.
The chain exhibits a virtual site (represented in green) which  drains the excitonic population with a rate $\Gamma$.
This type of directed chain slightly differs from the dissipative chains shown in our work (see Fig.~\ref{fig:directed_net}). Here, the associated rate matrix $\tilde{\bold{K}}$ (with dimension $(L+2)\times (L+2)$)  is defined as 
  \begin{equation} \label{eq:shape_K_tilde}
     \bold{\tilde{K}} =  
     \begin{bmatrix} 
      0 & \Gamma  &  0  &    &      & \\
      0  & -\Gamma - k^\text{right}_1 & k^\text{left}_2  &    &      &   \\
           &k^\text{right}_1          & - k^\text{right}_2- k^\text{left}_2 & k^\text{left}_3 &    &  \\
         &                           & k^\text{right}_2 & \ldots &   \ldots &   \\
         &                          &  &  \ldots &   \ldots & k^\text{left}_{L+1}\\
          &                          &  &   &      k^\text{right}_{L}& - k^\text{left}_{L+1}\\
     \end{bmatrix}.  
 \end{equation}
It encodes the irreversible excitonic transfer towards the virtual site (with index $0$ in Fig.~\ref{fig:WTDF}).

Based on the rate matrix $\bold{\widetilde{K}}$, one can introduce the matrix $\bold{Q (z)}$ (size $(L+2) \times (L+2)$) whose elements $Q_{ij} (z)$ (expressed in the Laplace domain) encode the so-called ``waiting time distribution functions'' as
\begin{equation}
    Q_{ij}(z) = [ \bold{\widetilde{K}}  \frac{1}{\bold{I} z + \bold{\widetilde{K}_D}}]_{ij}. 
    \label{eq:Qij}
\end{equation}
As explained in Ref.~\cite{silbey08}, this element is the Laplace transform of the time-dependent transition probability for a walker to travel from site $j$ to site $i$ on the network.
Here, $\bold{\widetilde{K}_D}$ refers to the diagonal matrix associated to the whole rate matrix $\bold{\widetilde{K}}$, $\bold{I}$ is the identity matrix and $z$ denotes the Laplace variable.
From this matrix, one could determine the associated ``probability distribution functions'' noted $\phi_{i}(z)$  (still in Laplace domain). 
% The latter describe the probability for the exciton  to occupy the site $L$ ===== depends on the adjacent state probability distribution functions and transitions probabilities such as : 
It describes the probability for the exciton  to occupy a given site $i$ assuming that it started from the rightmost site on the network shown in Fig.~\ref{fig:WTDF}. 
These quantities are defined via recurrence relations (see Ref.~\cite{silbey08}) which read
\begin{equation}
    \phi_{i}(z) = \phi_{i+1}(z) Q_{i+1,i} (z) + \phi_{i-1}(z) Q_{i-1,i} (z). 
    \label{eq:reso_rate}
\end{equation}

Within this context, the key quantity one needs to compute is 
$\phi_{L+1}(z)$. It represents the Laplace transform of the first passage time distribution for reaching the virtual site (indexed 0 in Fig.~\ref{fig:WTDF}) provided that the process starts from the rightmost site. It can be determined recursively using Eq.~(\ref{eq:reso_rate}), while verifying the boundary conditions $\phi_{L+1}(z) = \phi_{L}(z) Q_{L,L+1} (z)$ and $\phi_{0}(z) = 1 $. 
%By doing so, one can then determine the probability for the exciton to reach the virtual site depending only on the rate constants coded in $\tilde{\bold{K}}$. 
%
Once $\phi_{L+1}(z)$ is defined, one can then obtain directly the associated excitonic absorption time $\Bar{\tau}$ by evaluating the following first order derivative~\cite{gopich03,gopich06,silbey09,singh2022}:
\begin{equation}
   \Bar{\tau} = - \left.\frac{\delta \phi_{L+1}(z)}{\delta z} \right|_{z=0}.
\end{equation}
Applying this method to the directed chain  shown in Fig.~\ref{fig:WTDF}, one recovers the analytical form of the excitonic absorption time given in Eq.~(\ref{eq:def_rates}).

%----------------------------------------------------------------------------------------
%	REFERENCE LIST
%----------------------------------------------------------------------------------------
\phantomsection
\bibliography{biblio}

%----------------------------------------------------------------------------------------

\end{document}